\documentclass[12pt,preprint]{aastex}

\usepackage{emulateapj5}

\renewcommand\>{{\rangle}}
\newcommand\simgt{\lower.5ex\hbox{$\; \buildrel > \over \sim \;$}}
\newcommand\msun{{\rm\,M_\odot}}

\newcommand\au{{\rm\,AU}}

\newcommand\cm{{\rm\,cm}}

\newcommand\K{{\rm\,K}}
\newcommand\<{{\langle}}
\newcommand\LB{{\bf A}}
\newcommand\LC{{\bf B}}
\newcommand\LD{{\bf C}}

\def\myputfigure#1#2#3#4#5%
{\vskip#5pt\makebox[0pt]{\hskip#2in
\includegraphics[width=#3\textwidth]{#1}}\vskip#4pt\hfill}

\shortauthors{Gammie et al.}
\shorttitle{Clump Shapes}

\begin{document}

\title{Analysis of Clumps in Molecular Cloud Models: \\ 
Mass Spectrum, Shapes, Alignment and Rotation}

\author{Charles F. Gammie and Yen-Ting Lin}

\affil{Center for Theoretical Astrophysics, University of Illinois \\
1002 W. Green St., Urbana, IL 61801, USA; gammie@uiuc.edu, ylin2@astro.uiuc.edu}

\author{James M. Stone and Eve C. Ostriker}

\affil{Astronomy Department, University of Maryland, \\
College Park, MD 20742 USA.; jstone@astro.umd.edu, ostriker@astro.umd.edu}

\begin{abstract}

Observations reveal concentrations of molecular line emission on the
sky, called ``clumps,'' in dense, star-forming molecular clouds.  These
clumps are believed to be the eventual sites of star formation.  We
study the three-dimensional analogs of clumps using a set of
self-consistent, time-dependent numerical models of molecular clouds.
The models follow the decay of initially supersonic turbulence in an
isothermal, self-gravitating, magnetized fluid.  We find the following.
(1) Clumps are intrinsically triaxial. This explains the observed
deficit of clumps with a projected axis ratio near unity, and the
apparent prolateness of clumps.  (2) Simulated clump axes are not
strongly aligned with the mean magnetic field within clumps, nor with
the large-scale mean fields. This is in agreement with observations.
(3) The clump mass spectrum has a high-mass slope that is consistent
with the Salpeter value.  There is a low-mass break in the slope at
$\sim 0.5 \msun$, although this may depend on model parameters including
numerical resolution.  (4) The typical specific spin angular momentum of
clumps is $4 \times 10^{22}\, {\rm cm^2\,s^{-1}}$. This is larger than
the median specific angular momentum of binary stars.  Scaling arguments
suggest that higher resolution simulations may soon be able to resolve
the scales at which the angular momentum of binary stars is determined.

\end{abstract}


\section{Introduction}

Star forming molecular clouds are observed to be highly inhomogeneous in
molecular line emission and continuum absorption studies.  They contain
relatively empty voids as well as dense ``clumps'' and denser ``cores''
within these clumps.  One expects that stars will form in the densest
parts of molecular clouds, and clumps are indeed correlated with young
stars \citep{wdb94,pl97}.  Clumps (or cores) therefore appear to be the
immediate precursors of clusters, small groups, or individual stars in
the interstellar medium; they constitute the initial conditions for star
formation.  

Theoretical models of diffuse precursors to stars were, until recently,
confined to static or quasi-static equilibrium models.  Mouschovias
(1976) constructed numerical models of static, axisymmetric,
magnetically supported, self-gravitating equilibria (see also:
\citealt{bame96, ls96, gs93, bes89, tin88a, tin88b, mera85}).  These
models do not include the effects of turbulence.  Quasi-static models,
in which the effects of turbulence are included as an effective
pressure, fixed by an assumed equation of state, have been constructed
by \citet{ls89} (see also \citealt{myf92, mcpu96, cm00}).

Both the static and quasi-static approach have weaknesses.  The static,
magnetically supported models predict oblate cores, flattened
perpendicular to the mean magnetic field (but see \citealt{fp00}).
Observations yield shapes apparently more consistent with prolate clumps
\citep{mfgb91,ryd96}, and little correlation between field
direction and clump shape (\citealt{hgmz}).  They also do not include the
effects of turbulent velocities, which are directly observed through
their contribution to the linewidth of molecular line transitions 
even at small spatial scales (e.g. \citealt{goo98}).  

The quasi-static models, on the other hand, account for turbulent
velocities by invoking a steady turbulent pressure.  It not immediately
apparent that supersonic turbulence like that observed in molecular
clouds would act to prevent collapse, since the turbulence itself leads
directly to compressions that might accelerate collapse.  Recent
numerical simulations show, furthermore, that supersonic magnetized
turbulence dissipates in a dynamical time (\citealt{sog98, mac99}).  If
these simulations correctly represent conditions in molecular clouds
then even an approximate steady state cannot be achieved unless momentum
is continually injected into the cloud.

Three-dimensional (3D), dynamic numerical models of molecular clouds
have recently become practical (e.g., \citealt{vsopg98} and the references
therein; \citealt{sog98, mac99, pan99, pic00, osg01}).  They do not impose
a high degree of symmetry on the cloud and allow a largely
self-consistent treatment of turbulence, although due to numerical
diffusion they have a much lower effective Reynolds numbers than real
interstellar clouds.  In this paper we will study the density
concentrations that arise self-consistently from self-gravitating
turbulence in a set of numerical models.

Before we proceed, it is worth discussing the limitations of the model and the
approximations that we use.  Our models integrate the equations of isothermal,
compressible, self-gravitating, ideal magnetohydrodynamics (MHD).  The isothermal approximation is
used because realistic, time-dependent heating and cooling that treats the
transfer of line radiation out of the cloud is, at present, out of numerical
reach (but see \citealt{jpn01}).  The model is fully self-gravitating, however,
we use periodic boundary conditions in our solution of the Poisson
equation.  This
distorts the gravitational field from what it would be in an isolated cloud by
introducing additional tidal forces (although real clouds are not isolated,
tidal forces are smaller).  The numerical model is ``ideal'' in the
sense that we do not include any explicit physical diffusion coefficients.  All
dissipation occurs in shocks, where it is captured by an artificial viscosity,
or via numerical diffusion at the grid scale.  Thus we do not model physical
diffusion processes such as ion-neutral drift.  This is almost certainly
important in the dense regions of clouds, but an accurate treatment of the
ionization, particularly UV ionization, that would motivate dynamical studies
of non-ideal clouds is also numerically out of reach.  

This paper is organized as follows:  In \S 2 we give our definition of a
``clump.'' In \S 3 we describe our numerical methods and tests and give a
qualitative and quantitative description of the evolution of the runs.  The
clump mass spectrum is described in \S 4.  The distribution of clump shapes in
described in \S 5, the orientation of the clumps with respect to the local and
global magnetic field is described in \S 6, and a discussion on the clump spin
angular momentum is given in \S7.  A summary and comparison with earlier
work is given in \S 8.

\section{Clumps: Definitions}

Our first task is to define a ``clump.'' Any definition is somewhat
arbitrary, much as the designation ``mountain'' \footnote{ Merriam-Webster:
``a landmass that projects conspicuously above its surroundings and is
higher than a hill.''} is arbitrary in topography.  The definition
should be: (1) physically motivated; (2) related to definitions used by
observers in molecular line studies of clouds; (3) simple to describe; (4)
easy to apply.  Ultimately we are motivated by the desire to characterize
those dense regions in molecular clouds that turn into stars.

Myers (1999) adopted the following definitions: a region of size $\simgt 10$ pc
over which mean hydrogen number density $n_{H_2} \sim 10^2 \,{\rm cm}^{-3}$ is
called a ``complex'', a region of size $\sim 1$ pc over which $n_{H_2} \sim
10^3 \,{\rm cm}^{-3}$ is termed a ``clump'' and a ``core'' refers to a region
of size $\sim 0.1$ pc over which $n_{H_2} \sim 10^4 \,{\rm cm}^{-3}$. A complex
is well sampled by the CO $J = 1 \to 0$ transition, a clump can be traced by
the $J = 1 \to 0$ line of $^{13}$CO while a core is traced by the $(J,K) =
(1,1)$ line of ${\rm NH}_{3}$. The choice of $10^4 \,{\rm cm^{-3}}$ as the
fiducial density has several advantages, e.g., some easily observable molecular
lines have effective critical density of this order. 

There are two frequently-applied algorithms for identifying clumps in
molecular line data.  Each begins with intensity data in two space plus one
velocity coordinates ($l-b-v$ space).  The first algorithm, GAUSSCLUMPS,
uses a fitting procedure that removes objects with a constrained gaussian
shape from the data (e.g. \citealt{sgu90});  it is analogous to the CLEAN
algorithm used in radio interferometry.  The second algorithm, CLUMPFIND,
identifies intensity maxima in $l-b-v$ space (\citealt{wdb94}), then assigns
each ``pixel'' (i.e., each channel in a spectrum) to be a member of the
clump associated with the nearest intensity maximum.

These algorithms all implicitly make the assumption that the velocity
coordinate $v$ is somehow related to the line of sight distance $z$, i.e.  that
widely separated material that happens to lie at the same velocity is not
confused in projection.  As we have previously shown using simulated data sets
(\citealt{osg01}; see also \citealt{pic00, bpm02}), 
this assumption may frequently {\it not} be satisfied for
moderate-density condensations, although for rarer high-density peaks it
probably is.  With a suitable source function and radiative transfer scheme, it
is possible to generated simulated $l-b-v$ data cubes for molecular transitions
with varying critical densities, and thus to ascertain exactly when $v$ is
expected to be a valid surrogate for $z$.  We defer consideration of this
interesting and important problem to a later publication.

For the present work, we take advantage of our direct access to all
three spatial coordinates in simulated data to study clumps in the
unprojected density field.  We associate clumps with spatial maxima in
the density in much the same way that CLUMPFIND associates clumps with
maxima in the intensity.  Our algorithm operates as follows.  Find all
density maxima ${\bf r}_i$.  Then associate every zone ${\bf r}$ that
lies above a threshold density $\rho_t$ with the nearest ${\bf r}_i$.
Each ${\bf r}_i$, together with the surrounding material that lies above
the threshold density, is then a clump.

This algorithm is easy to implement and explain.  It has one parameter
$\rho_t$.  A potential problem is that the density may contain small
variation near the density maxima (are the Grand Tetons one mountain, or
three?).  If these variations occur on the grid scale, they may be purely
numerical in origin.  We can test the degree to which these variations
split up what would otherwise be single clumps by smoothing the density and
picking ${\bf r}_i$ to be maxima of the smoothed density.  This introduces
one additional parameter, the smoothing length $\Delta$.

What is the relationship of the 3D clumps described here with those seen
by observers, in projection?  As mentioned above, preliminary work
(\citealt{osg01}) indicates that projection effects can be important.  In
each section below we will apply the same clump analysis to both the
column density and the density as a way of evaluating the importance of
projection effects.

\section{Numerical Method, Tests, and Run Evolution}

We evolve the equations of isothermal, self-gravitating, compressible,
ideal MHD using a numerical method based on the ZEUS code
\citep{sn92a,sn92b}.  It is a time-explicit, operator-split finite
difference algorithm on a staggered mesh.  Density and internal energy are
zone-centered, while velocity components live on zone faces.  The magnetic
portion of the code uses the ``method of characteristics'' to evolve the
transverse components of the magnetic and velocity field in a manner that
assures the accurate propagation of Alfv\'en waves.  The magnetic field is
represented by fluxes through zone faces; these fluxes are evolved using
``constrained transport,'' \citep{eh88} which preserves the constraint
$\nabla\cdot{\bf B} = 0$ to machine precision.

The gravitational acceleration is obtained by differencing the
gravitational potential.  The potential is calculated using a Fourier
transform technique.  The gravitational kernel is defined by $\phi_{\bf k} = 4\pi
G\rho_{\bf k}/ (\sum_i (2 \cos (k_i \Delta x_i) - 2)/(\Delta x_i)^2)$.
This kernel ensures that the natural finite difference form of the
Poisson equation is satisfied.  Notice that in a periodic domain the
Poisson equation becomes
\begin{equation}
\nabla^2 \phi = 4\pi G (\rho - \bar{\rho}).
\end{equation}
Evidently the periodic boundary conditions alter the potential from what
it would be in an isolated cloud.  The difference is most pronounced
away from density maxima.  In regions with $\rho \gg \bar{\rho}$ the
potential is less affected by the periodic boundary conditions.  This
can be verified by explicit construction of sheetlike equilibria, which
we have done.

We use an isothermal equation of state $p = c_s^2\rho$, $c_s^2 = const.$
This models the effect of efficient heating and cooling, and is
appropriate when the cooling time is much shorter than the dynamical
time.

Numerical convergence is a serious concern for the calculations
presented here, particularly for the case of clumps that are only a few
zones in size.  The self-gravitating isothermal MHD equations will,
under a broad range of circumstances, evolve toward density
singularities (i.e.  stars) that are unresolved.  Many aspects of
``protostellar'' turbulence, however, may be insensitive to the internal
structure of small, high-density regions.  Certainly this is the
assumption on which our numerical investigation is undertaken.  At a
minimum, however, we must only measure quantities on well-resolved
scales.

\subsection{Initial and Boundary Conditions}

This paper considers a set of three numerical models which differ in
magnetic field strength.  Several aspects of evolution and structure for
the same three models have previously been investigated by \citet{osg01}.
Each model has a numerical resolution of $256^3$ zones.

The models are evolved in a periodic cubic domain of length $L$.  The
initial density and magnetic field are uniform.  The initial velocity
field, $\delta {\bf v}$,  is chosen to be a divergence-free Gaussian
random field with power spectrum $v_k^2 \sim k^{-4}$.  This slope is
compatible with Larson's law; see, e.g., \citet{mga99}.  The
perturbations have $E_{w} \equiv {1\over{2}}\int d^3 x\,(\rho\,\delta
v^2 + \delta B^2/4\pi) = 100\,\rho\,L^3 c_s^2$, although initially
$\delta B = 0$.  No further perturbations are added over the course of
the evolution; these are {\it decay} simulations.

An important parameter of the model is the strength of selfgravity.
At fixed physical density and sound speed, this strength is directly
related to the physical size of the model $L$.  A suitable
nondimensional parametrization is in terms of the ``Jeans number'' $n_J
\equiv L/L_J$, where $L_J^2 = \pi c_s^2/(G \bar{\rho})$.  All models
discussed here have $n_J = 3$.

A second parameter is the strength of the mean magnetic field, $\< B_x \>$,
where $\<\>$ indicates a volume average.  $\< B_x\>$ cannot change over
time because of the periodic boundary conditions.  A sense of the dynamical
importance of the mean field may be had by first comparing $v_{A,x} \equiv
\< B_x \>/\sqrt{4\pi\bar{\rho}}$ to the sound speed $c_s$.  Our models have
$v_{A,x}/c_s = 1$ (``weak'' field), $\sqrt{10}$ (``moderate'' field), and $10$
(``strong'' field). The strong and moderate field cases are most consistent
with observations (e.g. \citealt{c99}; but see \citealt{pan99} for a different
perspective). 

One may also compare the mean field strength to that required to prevent
gravitational collapse: $\<B_x\> > 2\pi \bar{\rho} L \sqrt{G}$ (Tomisaka
1988a,b).  In terms of our dimensionless parameters $v_{A,x}/c_s$ and
$n_J$, this becomes $v_{A,x}/c_s > \pi n_J$.  Models which satisfy this
condition are {\it subcritical}, i.e., will never experience complete
gravitational collapse, while systems which fail to satisfy it are {\it
supercritical}.  \footnote{The mass-to-flux ratio in our models is always
constant.  Relaxing this constraint introduces free functions, rather than
a free parameter, in the initial conditions.  These degrees of freedom will
be explored in future models.  Notice that deciding whether a cloud is sub-
or super- critical requires knowledge of the field strength {\it and}
geometry.} In the present work, the strong field model is subcritical
and the other two models are supercritical.

\bigskip
\myputfigure{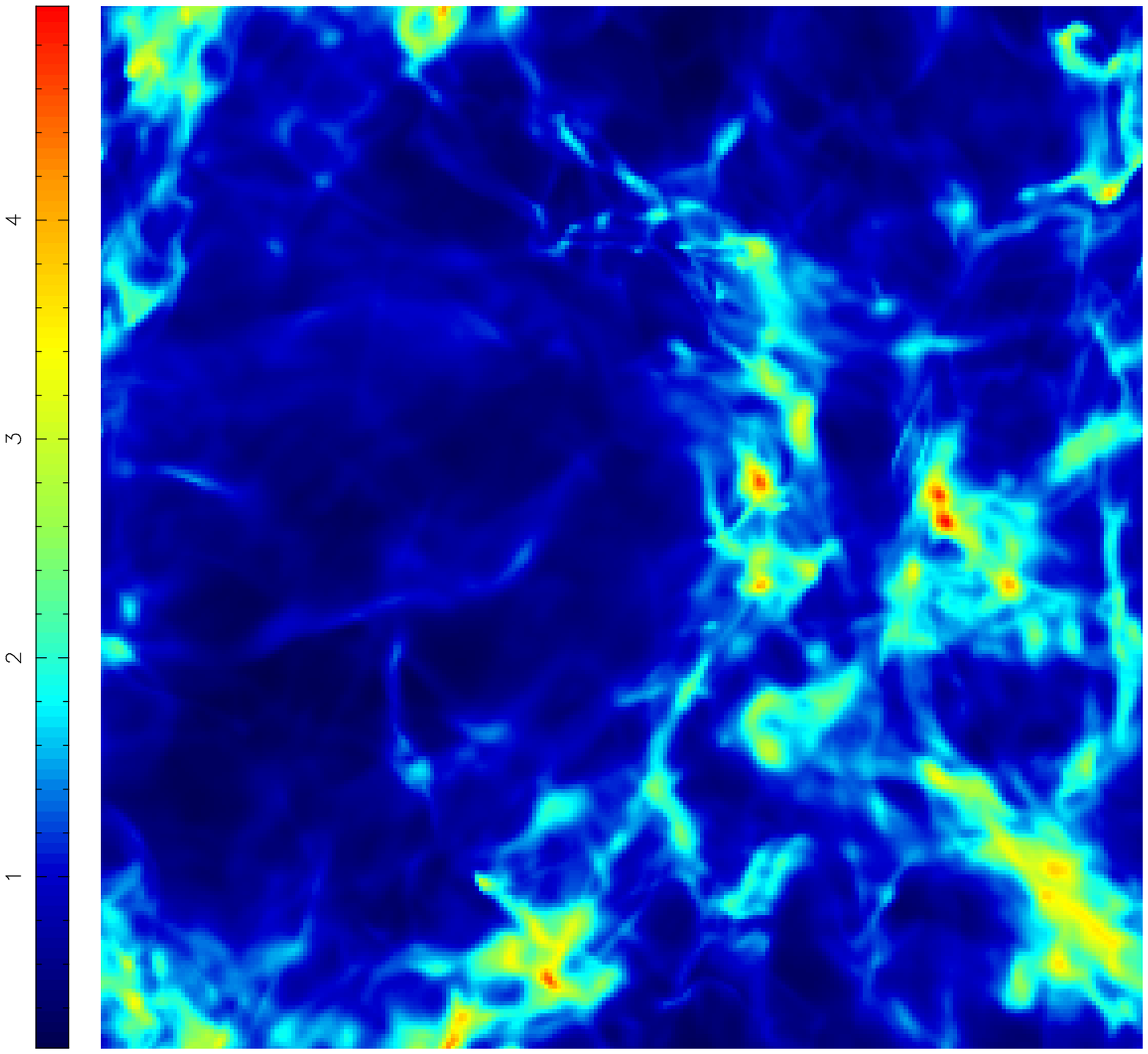}{3.2}{0.45}{-0}{-0}
\figcaption{
A color image of the surface density from snapshot \LC\ 9.  
The scale bar on the left shows column density relative to
the mean column density.
}

\subsection{Run Evolution}

Once the model evolution is started, the initially incompressive
velocity field distorts the initially uniform magnetic field.  Since the
magnetic field distortions are typically large, in the sense that
$\delta B^2/(8\pi) \gtrsim \rho c_s^2$, they couple strongly to
compressive motions, and density variations result.  Density variations
also result from exchange of power between the compressive and
incompressive parts of the velocity field; this would occur even in the
absence of a magnetic field.  Because the initial power spectrum of
velocities we have chosen has $v_\lambda$, the velocity variation at a
scale $\lambda$, varying as $\lambda^{1/2}$, and the timescale for a
density variation to form is $\lambda/v_\lambda$, the density variations
appear first at the smallest scales.  Within a dynamical time, $L/v_L$,
density variations have appeared at all scales and the flow is fully
turbulent.

Each run begins with a kinetic energy of $100\,\rho\,L^3 c_s^2$, which is
equivalent to an r.m.s. Mach number ($\mathcal{M} \equiv \sigma_v / c_s $)
of $\sqrt{200}$.  This kinetic energy is rapidly reduced by work done in
bending the mean field and in compressing the isothermal fluid.  It is also
dissipated in shocks.  The total ``wave energy'' (difference between the
energy and that of the initial state in the absence of perturbations)
dissipates on a timescale of order $L/v_L$, or $0.07 L/c_s$.  Figure 1
shows a color-coded image of the surface density in the middle of one
run (\LC); here black is low density and red is high density.

In this paper we study a set of three identical experiments that
differ only in the strength of the mean field.  These are designated
runs \LB\,(strong field, $\beta \equiv c_s^2/v_{A,x}^2 = 0.01$),
\LC\,(moderate field, $\beta = 0.1$), and \LD\,(weak field, $\beta = 1$).
Model \LB\ is subcritical, and models \LC\ and \LD\ are supercritical.
Runs \LC\ and \LD\ undergo gravitational collapse in the sense that at
some locations the density becomes very large.  In run \LB\ the peak
density rises to about $100 \bar{\rho}$ after the simulation begins and
then drops to only $40 \bar{\rho}$ at the end of simulation. In
contrast, the peak density rises almost monotonically in runs \LC\ and
\LD, and at final stages, reaches $7.9 \times 10^4 \bar{\rho}$ (\LC) and
$4.1 \times 10^4 \bar{\rho}$ (\LD).  Note that this density violates the
Truelove criterion 
\footnote{This criterion checks if the size of the Jeans length is of
order a zone size or smaller.  The instantaneous Jeans length is
$\lambda_J^2 = \pi c_s^2/(G \rho).$ Since $c_s$ is constant in space and
time, we can rewrite the Truelove criterion as $\rho <
{1\over{\mathcal{T}^2}} {c_s^2 N^2 \pi \over{G L^2}}$; here $N$ is the
number of resolution elements along each axis, $L$ is the size of the
box, and $\mathcal{T}$, the ``Truelove number'' must be larger than
about $4$.  In simulations units this becomes $\rho < 455\bar{\rho}$.
Therefore we are not resolving the internal dynamics of the clump with
the highest peak density.  } 
\citep{tkmh97}.  At the end of \LC, the fraction of the mass lying at
densities greater than ($1,10,10^2,10^3,10^4$)$\bar{\rho}$ is
($75,20,4,2,1$)\%.  At the end of \LD, the fraction of the mass lying at
densities greater than ($1,10,10^2,10^3,10^4$)$\bar{\rho}$ is
($70,13,3,1.5,0.9$)\%.  Thus while high density, collapsed regions do
develop in both supercritical runs, only a small fraction of the mass is
in the highest density regions.  The fraction of mass we find in
collapsing regions is consistent with observations of the efficiency of
star formation in giant molecular clouds as a whole (e.g., the $2\%$
found by \citealt{myetal86}).

From each of the simulations we have picked a set of ``snapshots'' to
study intensively.  They are labeled according to the run and evolution
time (see Table 1).  Run \LC, the moderate field ($\beta = 0.1$) run, is
our ``fiducial'' run, i.e. the one we believe is most similar to a
molecular cloud.  Snapshot \LC9 is our fiducial snapshot, since it is
evolved enough that the velocity and magnetic field have had a chance to
relax, but not so evolved that all turbulence has decayed away.

\begin{table*}[htb]
\begin{center}
\caption{Runs and Run Snapshots}
\begin{tabular}{lccccccccc}

\hline
\hline

Run &  Snapshot & time      &               & peak & & & & \\
Name & number   & $[L/c_s]$ & $v_{A,x}/c_s$ & density & $E_{kin}$ & $E_{\delta B}$ & $t_f$ & resolution \\
\hline
\LB  & 0  & 0   & 10 & 1.00               & 100. & 0.00 & 0.297 &  $256^3$\\
    & 3  & 0.03 &    & $1.06 \times 10^2$ & 38.4 & 32.7 & \\
    & 7  & 0.07 &    & 98.6               & 27.7 & 14.6 & \\
    & 19 & 0.19 &    & 38.2               & 12.2 & 4.00 & \\
    & 29 & 0.29 &    & 38.5               & 8.56 & 2.26 & \\
\hline
\LC  & 0  & 0    &  $\sqrt{10}$ & 1.00              & 100. & 0.00 & 0.195 & $256^3$ \\
    & 3  & 0.03  &              & $1.16 \times 10^2$ & 38.5 & 19.7 &   & \\
    & 4  & 0.04  &               & $1.30 \times 10^2$ & 28.7 & 20.3 &   & \\
    & 9  & 0.09  &              & $1.74 \times 10^2$ & 11.9 & 13.4 &   & \\
    & 11 & 0.11  &              & $3.19 \times 10^2$ & 10.3 & 10.2 &   & \\
    & 19 & 0.19  &              & $7.91 \times 10^4$ & 7.31 & 4.10 &   &  \\
\hline
\LD  & 0  & 0    &  1   & 1.00              & 100  & 0.00 & 0.154 & $256^3$ \\
    & 3  & 0.03 &       & $1.49 \times 10^2$ & 44.1 & 5.15 & & \\
    & 5  & 0.05 &       & $1.34 \times 10^2$ & 26.0 & 5.80 & & \\
    & 9  & 0.09 &       & $1.91 \times 10^2$ & 12.0 & 5.97 & & \\
    & 11 & 0.11 &       & $2.96 \times 10^2$ & 8.16 & 6.00 & & \\
    & 15 & 0.15 &       & $4.06 \times 10^4$ & 4.57 & 5.54 & & \\
\hline
\end{tabular}
\end{center}
\end{table*}

\subsection{Parameters}

We have to assign values to both the threshold density $\rho_t$ and the
smoothing length $\Delta$ (given in units of the grid spacing) to obtain
a population of ``clumps''.  In addition to these two parameters there
are two scaling parameters that must be chosen to give a physical scale
to the simulation (there are two rather than three because gravity sets
a lengthscale in the model).  We fix the mean number density $n_{H_2} =
100\cm^{-3}$ and the temperature $T = 10\K$, corresponding to
typical values found in molecular clouds.  With these scaling parameters
$\bar{\rho} = n_{H_2}\,\mu = 3.84 \times 10^{-22} {\rm g\,cm^{-3}}$ for
$\mu = 2.4\, m_p$, $c_s = \sqrt{kT/ \mu} = 0.19\, {\rm km\,s^{-1}}$, $L
= n_J\,c_s \sqrt{\pi/(G \bar{\rho})} = 6.45\, {\rm pc}$ (since $n_J =
3$).   Also the total mass ${\rm M_{tot}} = \bar{\rho} L^3 = 1532\, {\rm
M_{\odot}}$.

Once the scaling parameters have been chosen there is a natural choice
for the threshold density $\rho_t$ -- the density corresponding to the
critical density for CO($J=1 \to 0$) line, $\sim 10^3 \cm^{-3}$.
\footnote{In a real cloud, of course, radiative trapping permits regions
of density lower than the critical value to contribute significantly to
the observed emission, so the $^{12}$CO-emitting regions in a real cloud
might be somewhat larger than our defined clumps; clumps observed in
optically-thinner lines like $^{13}$CO or C$^{18}$O clumps may be closer
in size to the condensations defined by our density threshold.}  Since
$\bar{n}_{H_2} = 100 \cm^{-3}$, $\rho_t = 10\,\bar{\rho}$.

There is no equally simple way to set the smoothing parameter $\Delta$.
The purpose of smoothing the dataset before picking out the density
maxima is to avoid separation of closely spaced density peaks into
separate clumps;  one does not want to attach too much significance to
objects separated by only one grid zone.  This suggests fixing $\Delta
\sim 1$.  The number of clumps decreases as smoothing is increased,
however, so choosing $\Delta$ too large would result in the loss of most
clumps and most of the information contained in the dataset.  We have
experimented with $\Delta = 1,1.5,$ and $2$.  All these give similar
results.  All results reported here are for $\Delta = 1.5$ unless
otherwise noted.  To avoid making an arbitrary choice of this kind one
may alternatively consider an ensemble of clump sets defined by allowing
$\Delta$ to vary from $1$ to $N/2$, where $N$ is the number of zones.  
This is operationally similar to
the technique employed in \citet{osg01,ost02} (and also in the analysis of
large-scale cosmological structure).

\medskip
\myputfigure{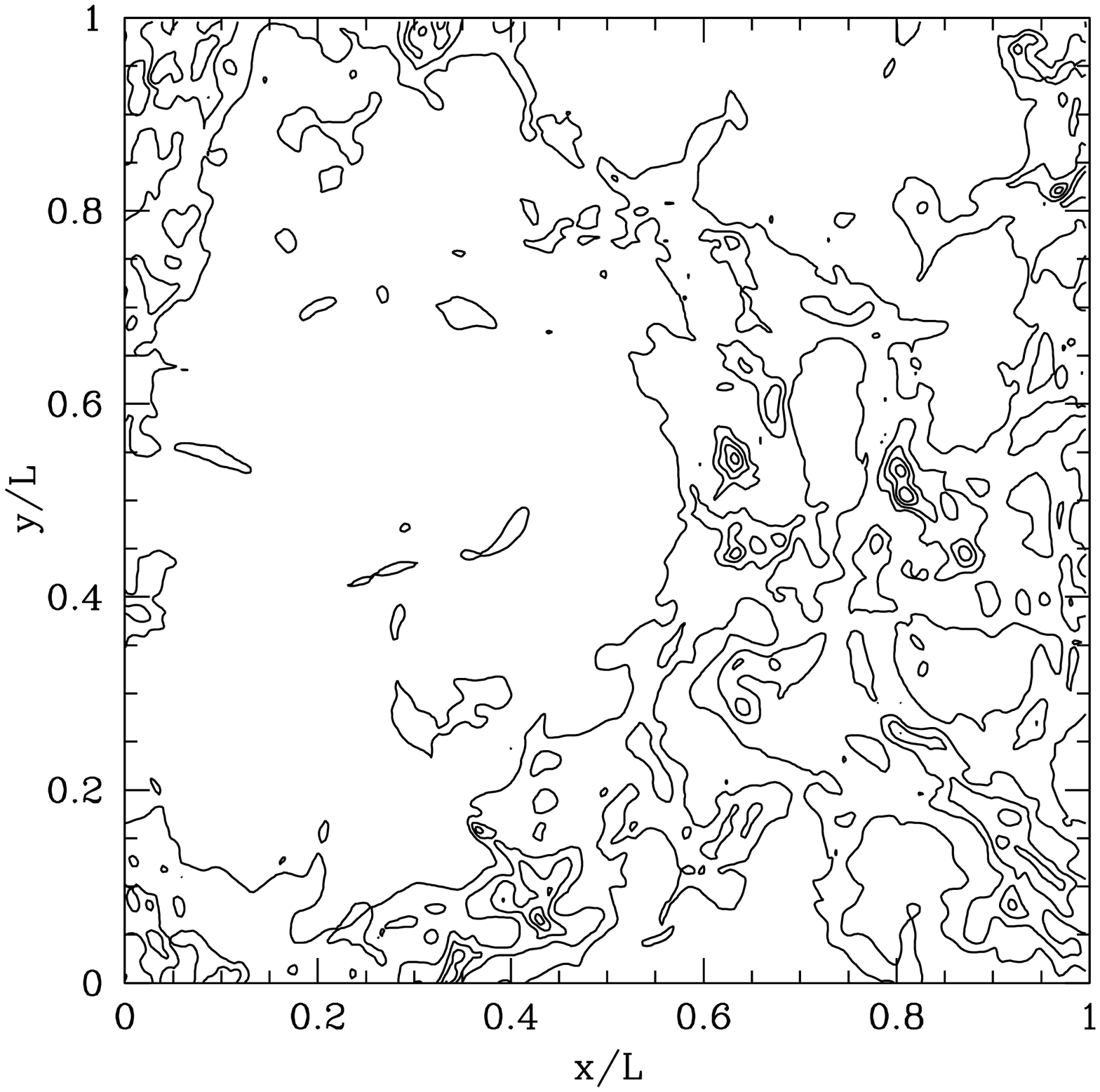}{3.2}{0.5}{-10}{-0}
\figcaption{
A contour plot of the surface density in snapshot \LC 9 from our
fiducial run.
}
\medskip

Are the clumps well resolved?  To decide this issue, we define an
effective radius $r_{eff}^2 \equiv \int \rho r^2/M_{cl}$, where the
integral is taken over the clump volume, $r$ is measured from the clump
center of mass, and $M_{cl}$ is the clump mass.  The distribution of
$r_{eff}$ depends on $\Delta$.  For $\Delta=1.5$ the clump size
distribution in snapshot \LC9 (our fiducial snapshot) is such that
$97\%$ of the clumps have an effective radius $r_{eff} \ge 3$ grid
zones, $78\%$ have $r_{eff} \ge 5$ grid zones, and $15\%$ have $r_{eff}
\ge 10$ grid zones.  Similar results are obtained for other snapshots.
Thus most clumps are (at this stage of the evolution) well resolved.

\myputfigure{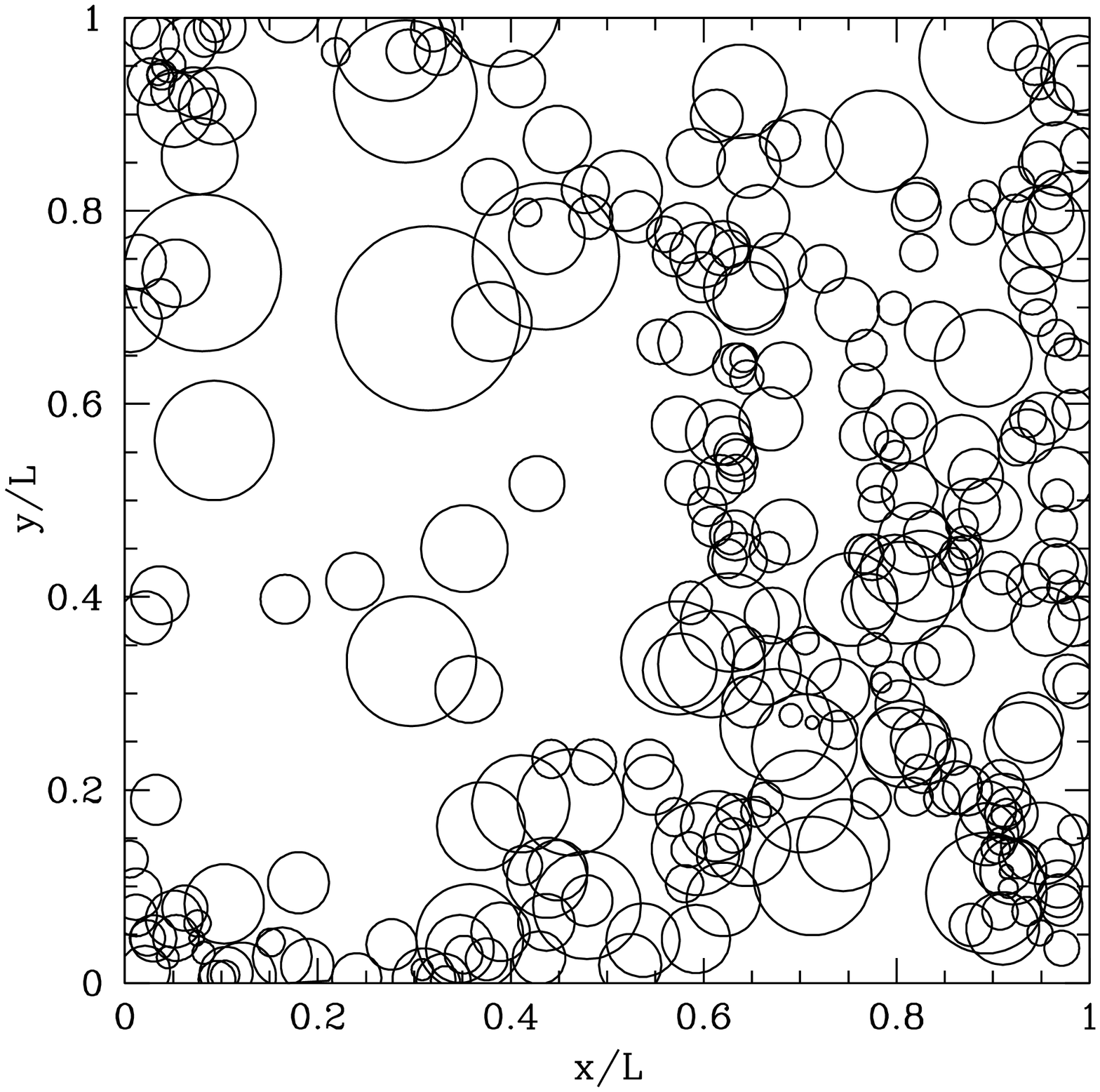}{3.2}{0.5}{-10}{-0}
\figcaption{
The location of 3D clumps (for definition see text) in snapshot \LC 9,
projected into the same plane as Figure 2.  The center of each circle 
is at the clump center of mass, and its radius is equal to the clump
effective radius.
}
\medskip

Are the 3D clumps identified with this set of parameters at all related
to 2D clumps (i.e. the clumps found by applying our clump finding
algorithm to column density maps)?  Figure 2 shows the column density
contours for a line of sight aligned with one of the simulation box
axes, and Figure 3 shows the positions of 3D clumps projected onto
the same plane.  The 3D clumps tend to be located in regions of large
column density, but the mass and volume fractions of 3D clumps and
2D clumps differ sharply.  The 3D clumps comprise $0.142\,{\rm M_{tot}}$
and $0.008\,{\rm L^3}$; in projection the 3D clumps cover $39\%$ of the
cloud.  The 2D clumps comprise $0.325\,{\rm M_{tot}}$ and $0.526\,{\rm
L^2}$.  The column density is thus a good tracer of the dense portions
of molecular clouds, but much of the contribution to each column may
come from regions with density below threshold or even from regions
spatially separated from 3D clumps (see \citealt{osg01}).

\section{Clump Mass Spectrum}

The simplest property of simulated clumps that we can study is their
mass spectrum.  We characterize the mass spectra by four numbers: the
maximum clump mass $M_{max}$, the minimum clump mass $M_{min}$, the mass
corresponding to the peak of the spectrum $M_{peak}$ (defined as the
maximum of $d N/d\ln M$ for data binned in $\ln M$) and the slope of the
high-mass wing, $x$.  The slope $x$ is determined by binning the clumps
in $\ln M$ and performing a linear fit to $d N/d\ln M \propto
M^{-(x-1)}$, weighted according to the square root of the number of
clumps in a bin in the range $M_{peak} < M < M_{max}$.  
Table 2 gives a summary of the mass spectrum characteristics for each of 
the snapshots.

In measuring $x$ our intent is simply to characterize the shape of the
mass spectrum in a way that is similar to that used by observers, not to
suggest that the underlying spectrum is a power law.  Other forms of the
mass spectrum may well provide a better fit.  Also notice that the
measurement of $x$ is noisy because of the small number of clumps in the
high mass wing.  It can change by $\pm 0.3$ depending on the choice of
bin size and location.

The minimum clump mass $M_{min}$ may be influenced by numerical
resolution.  We do not have a resolution study in hand and so we cannot
quantitatively evaluate this.  The minimum clump mass permitted by our
clump finding algorithm is the mass in a single zone that exceeds the
threshold density.  This is $(10/256^3) M_{tot} = 9.1 \times
10^{-4}\msun$, which is much smaller than our typical $M_{min}$.  This
is reassuring, but because the smallest clumps typically consist of a
small number of zones the reader is cautioned against placing too much
weight on our measurements of $M_{min}$.

\begin{table*}[htb]
\begin{center}
\caption{Statistics of Run Snapshots}
\begin{tabular}{lccccccccc}
\hline
\hline
Snapshot & no. & clump                     & $sg$ clump &     &  &  &  &  & \\
no.      & clumps & mass \tablenotemark{a} & mass \tablenotemark{b} & $x$\tablenotemark{c} & $x_{sg}$ & $x_{nsg}$ & $M_{max}/M_{tot}$ & $M_{peak}/M_{tot}$ & $M_{min}/M_{tot}$\\
\hline
\LB 3  & 304& 0.117& 0     &3.8 &   N/A & 3.8 & $2\times10^{-3}$ & $4\times10^{-4}$ & $4\times10^{-5}$\\
\LB 7  & 359& 0.131& 0     &2.7 &   N/A & 2.7 & $2\times10^{-3}$ & $3\times10^{-4}$ & $3\times10^{-5}$\\
\LB 19 & 82 & 0.033& 0     &1.6 &   N/A & 1.6 & $3\times10^{-3}$ & $2\times10^{-4}$ & $6\times10^{-5}$\\
\LB 29 & 32 & 0.017& 0     &0.5 &   N/A & 0.5 & $2\times10^{-3}$ & $4\times10^{-4}$ & $7\times10^{-5}$\\
\hline
\LC 3  & 285& 0.083& 0     &3.2 &   N/A & 3.2 & $2\times10^{-3}$ & $4\times10^{-4}$ & $2\times10^{-5}$\\
\LC 4  & 327& 0.096& 0.009 &2.7 &   N/A & 2.7 & $1\times10^{-3}$ & $2\times10^{-4}$ & $3\times10^{-5}$\\
\LC 9  & 300& 0.142& 0.138 &2.4 &   N/A & 2.5 & $5\times10^{-3}$ & $3\times10^{-4}$ & $6\times10^{-5}$\\
\LC 11 & 340& 0.163& 0.277 &2.3 & 2.0 & 2.5 & $6\times10^{-3}$ & $3\times10^{-4}$ & $4\times10^{-5}$\\
\LC 19 & 196& 0.207& 0.844 &2.0 & 1.9 & 1.9 & $1\times10^{-2}$ & $4\times10^{-4}$ & $7\times10^{-5}$\\
\hline
\LD 3  & 303& 0.113& 0.002 &3.0 &   N/A & 3.0 & $2\times10^{-3}$ & $4\times10^{-4}$ & $5\times10^{-5}$\\
\LD 5  & 298& 0.101& 0.127 &2.5 &   N/A & 2.5 & $2\times10^{-3}$ & $2\times10^{-4}$ & $4\times10^{-5}$\\
\LD 9  & 233& 0.110& 0.377 &2.2 & 1.6 & 2.1 & $4\times10^{-3}$ & $4\times10^{-4}$ & $6\times10^{-5}$\\
\LD 11 & 214& 0.129& 0.526 &2.3 & 1.8 & 2.0 & $7\times10^{-3}$ & $2\times10^{-4}$ & $6\times10^{-5}$\\
\LD 15 & 182& 0.141& 0.832 &1.6 & 1.2 & 2.1 & $2\times10^{-2}$ & $2\times10^{-4}$ & $3\times10^{-5}$\\
\hline
\end{tabular}
\tablenotetext{a}{Mass fraction in clumps.}
\tablenotetext{b}{Mass fraction in self-gravitating clumps.}
\tablenotetext{c}{The uncertainty in the mass spectrum slope $x$ is $\sim \pm 0.3$}

\end{center}
\end{table*}

Consider the mass spectrum of the fiducial snapshot \LC9, which is in
many respects typical (Figure 4, solid line).  A total of 300 clumps
were identified, accounting for $14.2$\% of the total mass and $0.871$\%
of the volume.  The slope of the spectrum is $x=2.35$ (it is an accident
that this is so close to the Salpeter slope for the stellar initial mass 
function [IMF], since
different values of the smoothing parameter $\Delta$ and different
binning schemes give values of $x$ that vary by as much as $\pm 0.3$).
The maximum mass $M_{max} = 5\times 10^{-3}\, {\rm M_{tot}}$, the
minimum mass $M_{min} = 6\times 10^{-5}\, {\rm M_{tot}}$, and the peak
mass $M_{peak} = 3 \times 10^{-4}\, {\rm M_{tot}}$.  In physical units
the peak mass is close to $0.5\, \msun$, and the minimum and maximum to
0.1 and 8 $\msun$, respectively.  Clearly this is an interesting range
of masses if clumps like these are the immediate precursors of star
formation, but it is important to remember the numerical limitations of
our study.  In particular we do not know how these values might vary
with initial conditions, Jeans number $n_J$, Mach number, or numerical
resolution.

We expect that some clumps are transients, the result of collisions
between oppositely moving streams of gas in which gravity plays no role,
while others are long lived and self-gravitating.  To measure the
importance of gravity for each clump we define
\begin{equation}
\alpha \equiv {E_g\over{2 E_k+3 E_p+E_b}},
\end{equation}
where $E_g = (1/2)\int (\rho - \bar{\rho})\,\phi\,d^3 r$ \footnote{This
is the gravitational contribution to the total, conserved energy under
periodic boundary conditions.}, $E_k = (1/2) \int \rho\,(v -
v_{cm})^2\,d^3 r$, $v_{cm}$ is the clump center-of-mass velocity; $E_p =
\int \rho\,c_s^2\,d^3 r$ is related to the internal energy and  $E_b =
\int B^2/(8\pi)\,d^3 r$ is the magnetic energy inside the clump; in each
case the integral is taken over the clump volume.  We somewhat
arbitrarily label clumps as self-gravitating if $\alpha \geq 0.5$ and
nonself-gravitating otherwise.  Notice that our $\alpha$ is slightly
different from the usual ``virial parameter'', which is proportional to
the ratio of gravitational potential energy to kinetic energy.  We will
analyze the kinematics of clumps by the virial theorem in greater detail
in a later paper (see, e.g., \citealt{zw90, mkzw92, bemk92, bpvss99}).

We show in Figure 4 the mass spectra for self-gravitating (dotted line)
and nonself-gravitating (dashed line) clumps in snapshot \LC9.  At this
stage only $4\%$ of the clumps (about $14\%$ of total clump mass) are
self-gravitating, and all self-gravitating clumps are in the
high-mass wing of the spectrum.  The majority of the clumps are
nonself-gravitating, and thus $x$ for all clumps and for the
nonself-gravitating subset are about the same.

\medskip
\myputfigure{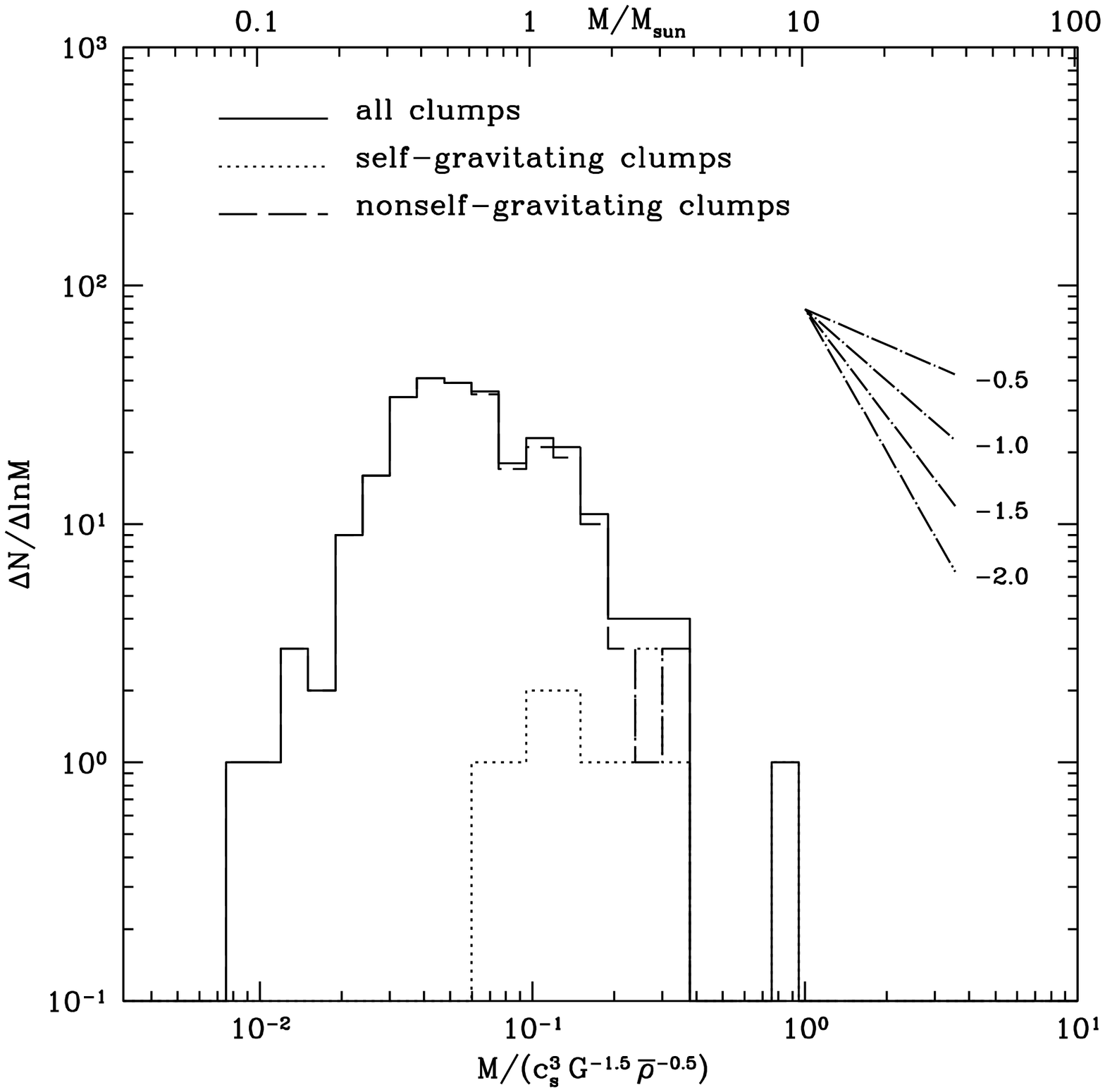}{3.2}{0.5}{-10}{-0}
\figcaption{
Mass spectra of self- and nonself-gravitating clumps and all clumps from
snapshot \LC 9.  The mass unit is $c_s^3\,G^{-1.5}\,\bar{\rho}^{-0.5}
\simeq 10.2 \msun$. 
}
\medskip

How does the clump mass spectrum evolve (see, e.g., \citealt{wibl98})?  In
Figure 5 we show the mass spectra from the fiducial, earliest and last
snapshots in run \LC.  Several features stand out: (1) $M_{max}$
increases with time.  From the first to last snapshot $M_{max}$ has
increased about an order of magnitude.  On the other hand, $M_{min}$
remains about the same.  (2) $M_{peak}$ remains about the same.  (3) The
number of clumps at peak (per bin) and the total clump number decrease
with time.  This is a result of clump agglomeration.  Notice that,
although in runs \LC\ and \LD\ the total clump number decreases with
time, the total mass contained in clumps increases.  Evidently clumps
also accrete mass from lower-density material.  (4) The slope of the
high-mass wing becomes ``shallower'', i.e., $x$ decreases:
$x(\LC3)=3.15,\, x(\LC9)=2.35, \,x(\LC19)=1.98$.  (5) The number of
self-gravitating clumps increases, especially in the high-mass wing: in
the fiducial snapshot there are only 13 self-gravitating clumps, but in
the last snapshot the entire high-mass wing is comprised of such clumps
(extending down to $M_{peak} \approx 0.6\,\msun$).

Does the field strength influence the mass spectrum?  Figure 6
shows mass spectra from snapshots \LB19, \LC9, and \LD9.  We have chosen
to compare these snapshots because all have similar sonic Mach numbers
$\mathcal{M}\approx 5$.
For the two supercritical runs (\LC\ and \LD) the mass spectra are
similar.  The subcritical run (\LB), however, is different: there are
essentially no self-gravitating clumps in any of the snapshots and the
mass spectrum changes in time in such a way that the number of clumps
{\it decreases} as the evolution progresses.

\myputfigure{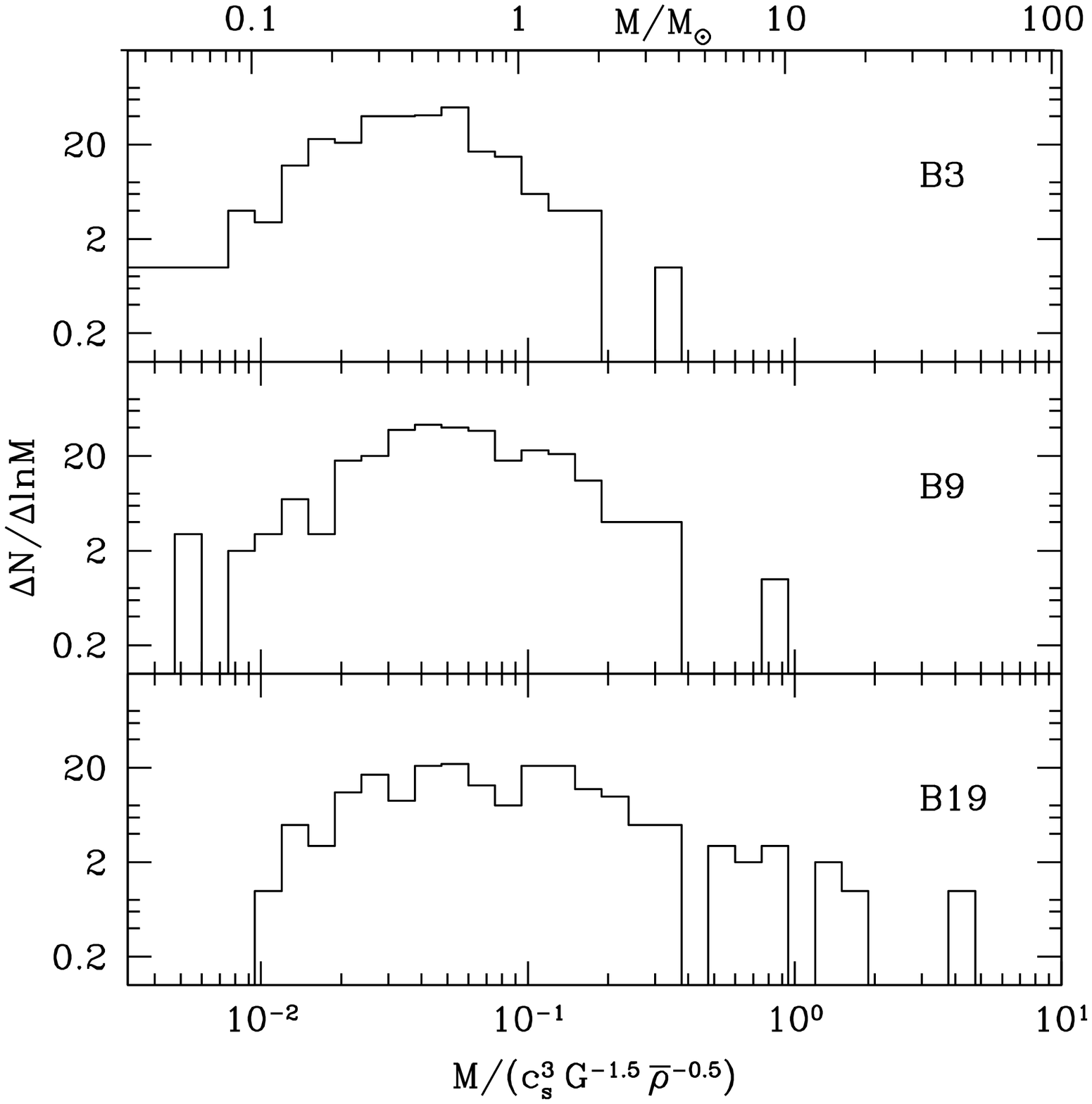}{3.2}{0.5}{-10}{-0}
\figcaption{
The evolution of the clump mass spectrum in the fiducial run \LC.
}
\medskip

Finally, do projection effects change the mass spectrum?  We have
examined the mass spectrum of clumps in column density maps made from
our simulations.  We use precisely the same clump definition for the
column density maps as in our 3D density fields, except that no
smoothing is applied ($\Delta = 0$, since the transformation to column
density is effectively a smoothing operation) and the threshold column
density is set to be $1.1$ times the mean column density.  We refer to
the resulting clumps as ``2D clumps''.  We find that (1) masses of 2D
clumps are larger than the 3D clumps (by about an order of magnitude)
and therefore, $M_{max}$ and $M_{p}$ are both greater than those in 3D
spectra.  (2) The slopes are generally slightly steeper than the 3D ones
and have a weak tendency to become shallower in time. For example,
$x(\LC3,2D)=3.26,\,x(\LC9,2D)=2.49, \,x(\LC19,2D)=2.50.$ The 2D and 3D
mass spectra for snapshot \LC9 are compared in Figure 7.

\section{Clump Shapes}

Clump shapes may be an indicator of the dynamical processes that govern
clump internal dynamics.  For example, early models of magnetized,
self-gravitating clouds (e.g. \citealt{mou76}) have oblate spheroidal
isodensity contours that are flattened in a plane perpendicular to the
mean magnetic field.  Observers have looked for evidence of such oblate
structures in molecular line maps of clouds.

\citet{mfgb91}(see also \citealt{dv87,f92,cus01}) measured
the projected axis ratios ($r$) of 16 dense cores in dark clouds and found
that $\<r\> = 0.5 - 0.6$.  They modeled ensembles of identical
spheroidal cores, either all prolate or all oblate, assuming that their
orientation in space is isotropic.  Their results showed that the
prolate ensemble can account for $\<r\>$ with reasonable
intrinsic axis ratio, while it requires {\it extremely} flattened oblate
cores to produce same projected axis ratios.  This suggests
that prolate objects are preferred, contrary to the prediction of
static, magnetized models.

Ryden (1996) obtained a tighter constraint on the shapes of clouds by
studying the full distribution of apparent axis ratios $f(r)$.  For
several samples of dense interstellar clouds, including Bok globules,
dense cores, and clumps, she mapped the estimated distribution $f(r)$
into a distribution of intrinsic axis ratios under the hypothesis that
the clouds were oblate or prolate spheroids.  She found that for most of
her cloud samples the data were inconsistent, or at best marginally
consistent, with the hypothesis that clouds are oblate spheroids because
of a lack of almost circular ($r \sim 1$) objects.  The samples were
consistent with the hypothesis that the clouds were prolate spheroids.

\medskip
\myputfigure{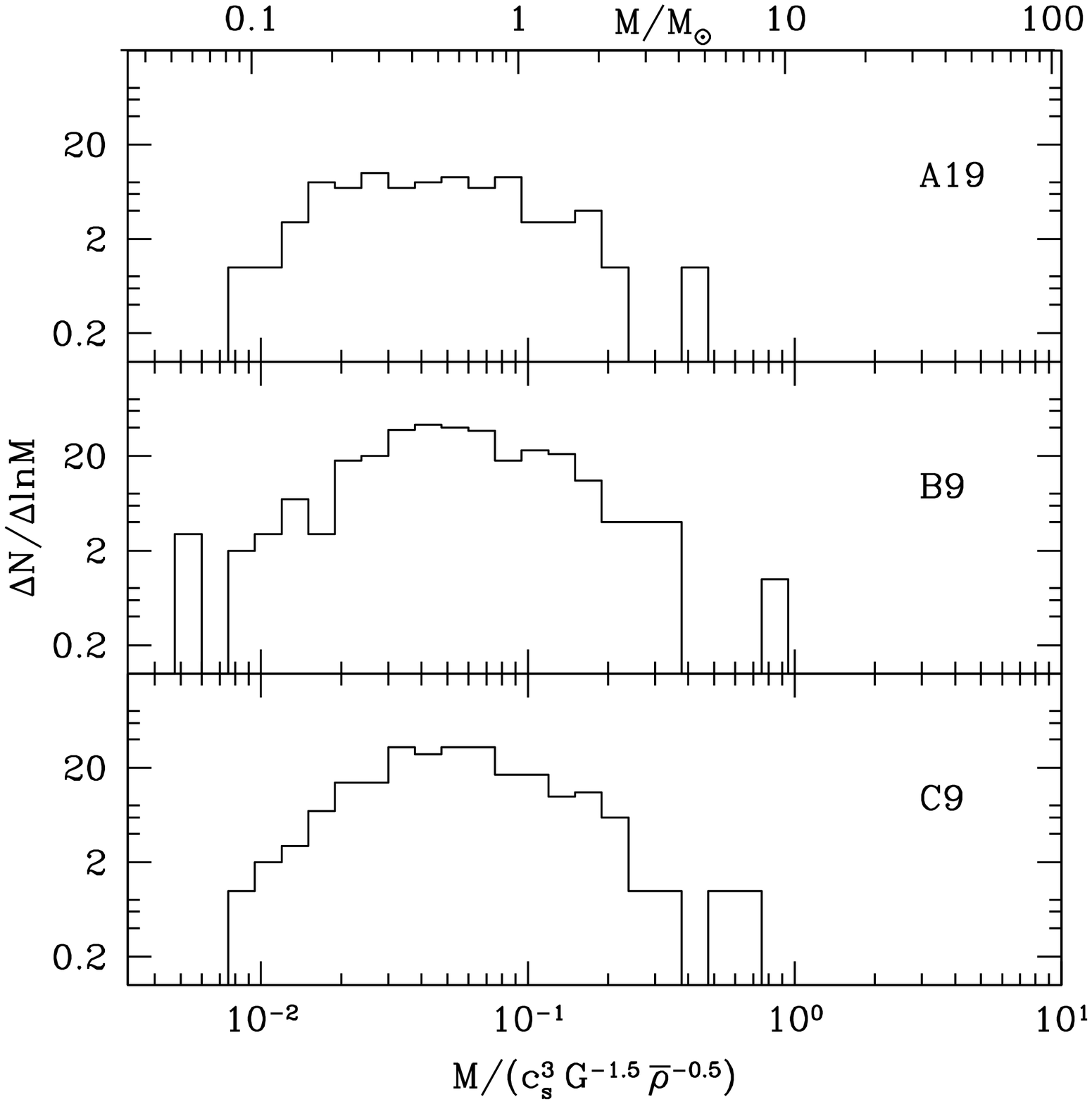}{3.2}{0.5}{-0}{-0}
\figcaption{
The dependence of clump mass spectrum on initial magnetic field strength
in snapshots from each of the three runs.  The snapshots are selected so
that the turbulent sonic Mach number is nearly constant.
}

\medskip
\myputfigure{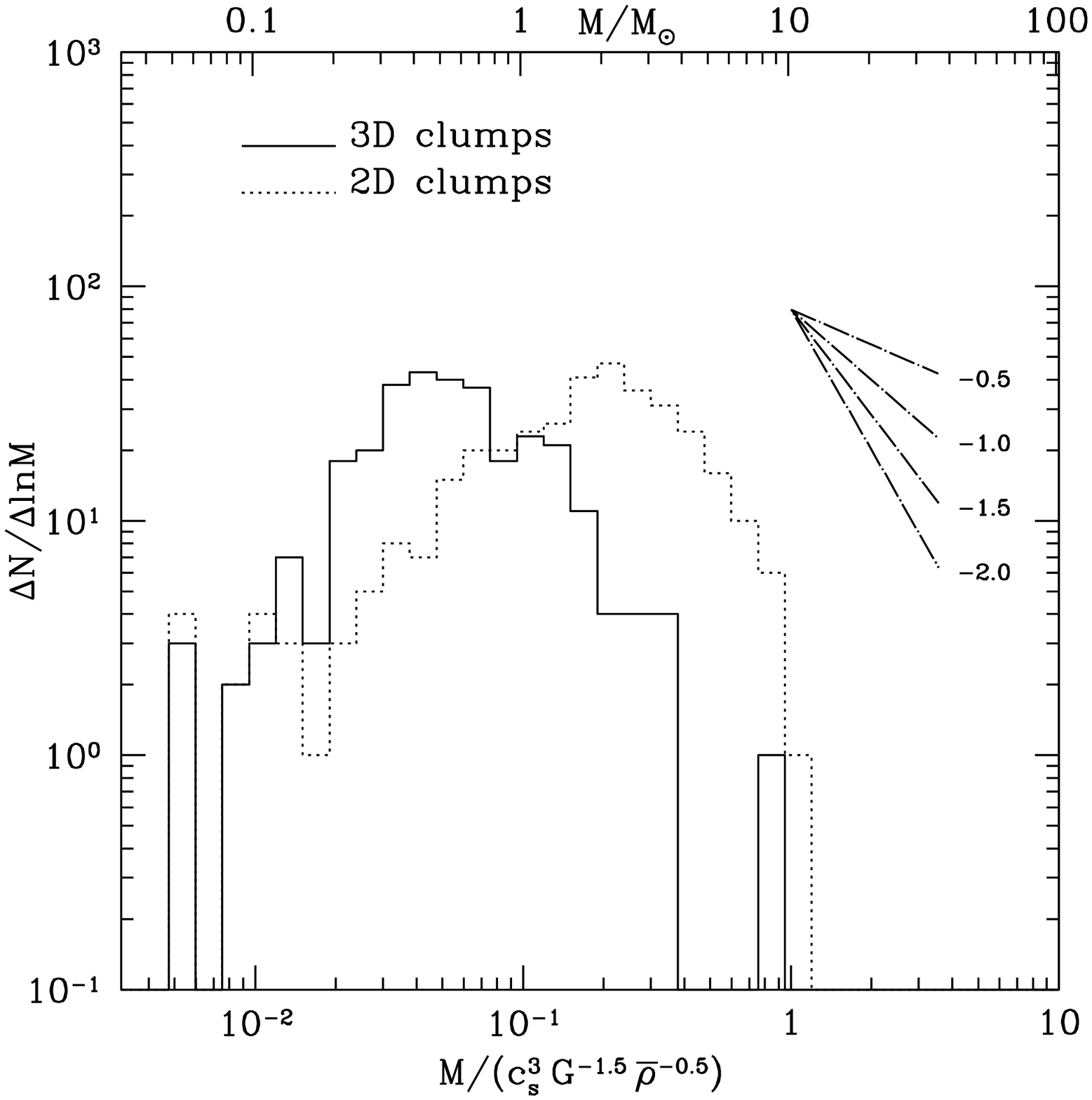}{3.2}{0.5}{-10}{-0}
\figcaption{
Comparison of clump mass spectra for clumps identified in three
dimensions (3D clumps) with those identified from column density maps in
\LC 9.  The differences arise in part from the overlapping of 3D clumps
in projection.
}

We have performed a similar analysis of a subset of the catalog of
clumps in the M17 SW cloud developed by \citet{sgu90} using their
GAUSSCLUMPS clump finder.  We have restricted attention to those clumps
with minor axis FWHM larger than $20''$ to minimize the effects of
rounding by the $13''$ FWHM beam, and we have subtracted the beam size
in quadrature from the axis lengths given in the paper.  Notice that
GAUSSCLUMPS uses a slightly different definition of clumps and principal
axis lengths from ours.  Given this sample of clump principal axis
lengths, it is possible to construct an estimate for the continuous
distribution $f(r)$ via the nonparametric kernel estimator used by Ryden
(we use a kernel width $h = 0.08$, on a sample of $41$ clumps).  The
result is shown in Figure 8 as a bold dashed line (Figure 9 shows the
same result applied to ``2D clumps''; see below).  The light dashed
lines show the uncertainties in $f(r)$ estimated via a bootstrap
resampling technique.  The mean axis ratio $\<r\> = 0.61$.  Recall that
$\<r\> > 0.5$ for oblate spheroids.

As for elliptical galaxies one can only infer the intrinsic shape of a
population of {\it spheroidal} clouds from a set of axis ratios.  If the
underlying population is triaxial, no such inference is possible, since
a mapping from a distribution with one axis ratios to that with two axis
ratios is degenerate.

\medskip
\myputfigure{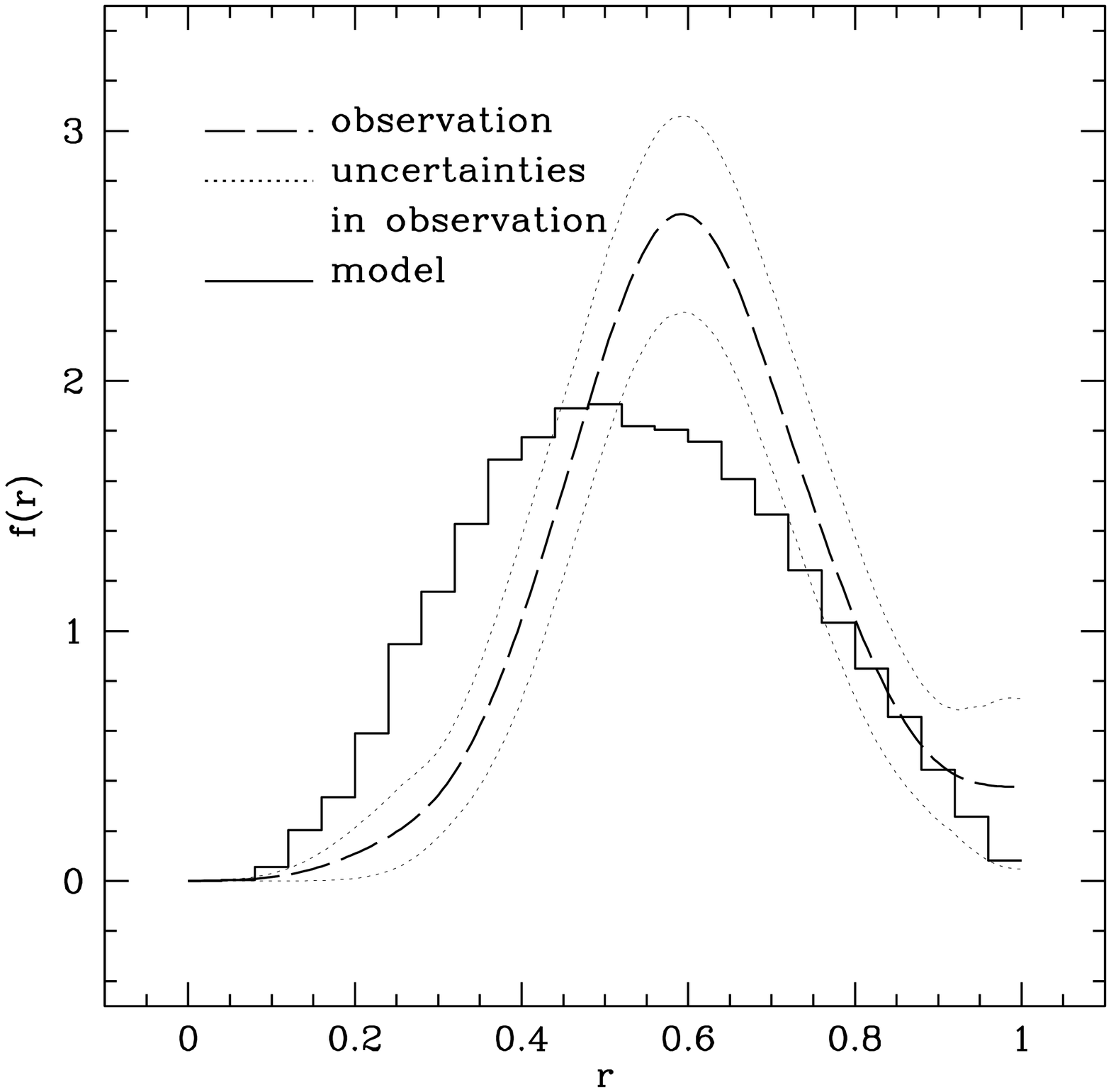}{3.2}{0.5}{-10}{-0}
\figcaption{
The distribution of apparent axis ratios of 3D clumps (histogram) in our
model and the distribution of apparent axis ratios in Stutzki \&
G\"usten's (1990) catalog of clumps from M17SW derived using Ryden's 
(1996) method.
}

\subsection{Clump Shape Definition}\label{SHAPDEF}

We characterize clump shapes using the eigenvalues of the moment of
inertia tensor:
\begin{equation}
I_{ij} \equiv \int d^3{\bf r}\,\rho\,x_i\,x_j.
\end{equation}
Here ${\bf x}$ is measured relative to the clump center of mass.
The eigenvalues are $M a^2 \ge M b^2 \ge M c^2$, where $M$ is the clump's
mass and $a,b,c$ are the principal axis lengths.  The corresponding
eigenvectors ${\bf e}_a, {\bf e}_b, {\bf e}_c$ are the principal axes.
The intrinsic shape can be characterized via the axis ratios
$\tilde{\beta} \equiv b/a$ and $\tilde{\gamma} \equiv c/a$ (note that
$\tilde{\beta}$ differs from the plasma $\beta$).

One advantage of using the moment of inertia tensor to measure clump
shape is that it can be related to observations, which are done in
two dimensions on the sky.  The projection of the moment of inertia
tensor is
\begin{equation}
P_{ij} \equiv \int d^2{\bf r}\,\Sigma\,X_i\,X_j.
\end{equation}
Here $X_{1,2}$ are Cartesian coordinates on the plane of the sky and
$\Sigma$ is the surface density or anything proportional to the surface
density, such as total flux in an optically thin line that traces the
density.  $P$ has eigenvalues $M p^2 \ge M q^2$.  The apparent shape is
characterized by the apparent axis ratio $r \equiv q/p$.  This is what
can be measured by an observer from column density maps of an isolated
clump.  

\medskip
\myputfigure{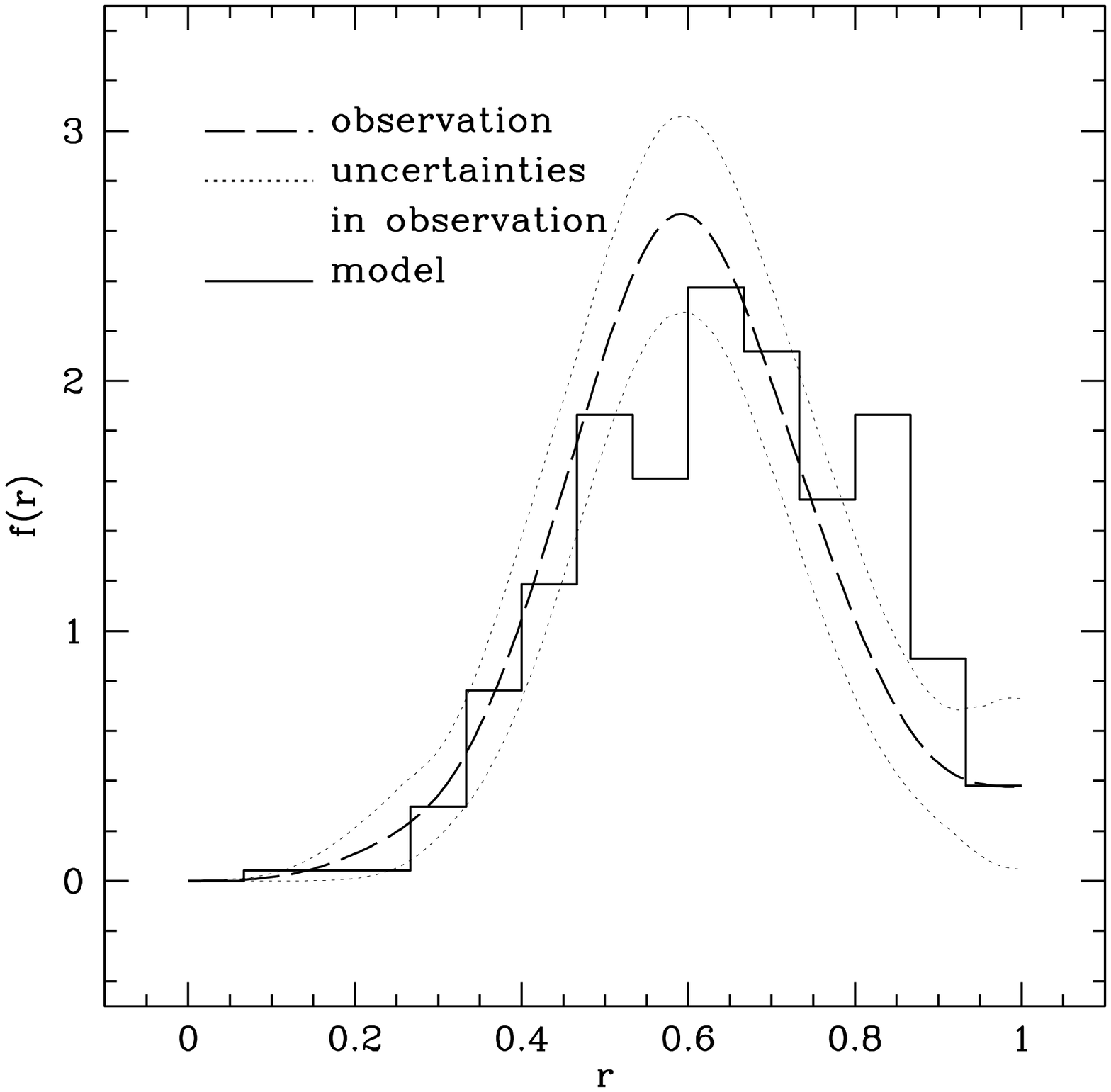}{3.2}{0.5}{-10}{-0}
\figcaption{
The distribution of axis ratios for 2D clumps (histogram) together with
the M17SW clump distribution from Stutzki \& G\"usten's (1990), as in
Figure 8.
}
\medskip

The apparent shape can be calculated from the eigenvalues $a,b,c$ of the
moment of inertia tensor without actually projecting the clump pixel
by pixel (this is computationally useful).  Let $\theta$ be the angle
between ${\bf e}_c$ and line-of-sight, and $\phi$ be the angle between
${\bf e}_a$ and ${\bf x}$, which points along the intersection of the
${\bf e}_a, {\bf e}_b$ plane with the plane of the sky.  Then the
apparent axis ratio is 
\begin{equation}
r = \left({P_{xx} + P_{yy} - \sqrt{P_{xx}^2 + 4 P_{xy}^2 -
        2 P_{xx} P_{yy} + P_{yy}^2}\over
        {P_{xx} + P_{yy} + \sqrt{P_{xx}^2 + 4 P_{xy}^2 -
        2 P_{xx} P_{yy} + P_{yy}^2}}}\right)^{1/2}
\end{equation}
where $P_{xx} = a^2 \sin^2\theta + b^2 \cos^2\theta\sin^2\phi + c^2
\cos^2\theta\cos^2\phi$, $P_{xy} = (b^2 - c^2) \cos\theta
\cos\phi\sin\phi$, $P_{yy} = b^2 \cos^2\phi + c^2 \sin^2\phi$.  

Finally, some terminology: the term ``projected clump'' as used here
refers to an object found by our clumpfinder in three spatial dimensions
(a 3D clump) and then projected as if it were completely isolated onto
the sky.  The term ``2D clump'' refers to an object found by our
clumpfinder in two spatial dimensions, after the entire density field
has been projected on the sky.  Were all 3D clumps well separated these
procedures would yield identical results.  However 3D clumps overlap in
projection and therefore the populations differ.

\medskip
\myputfigure{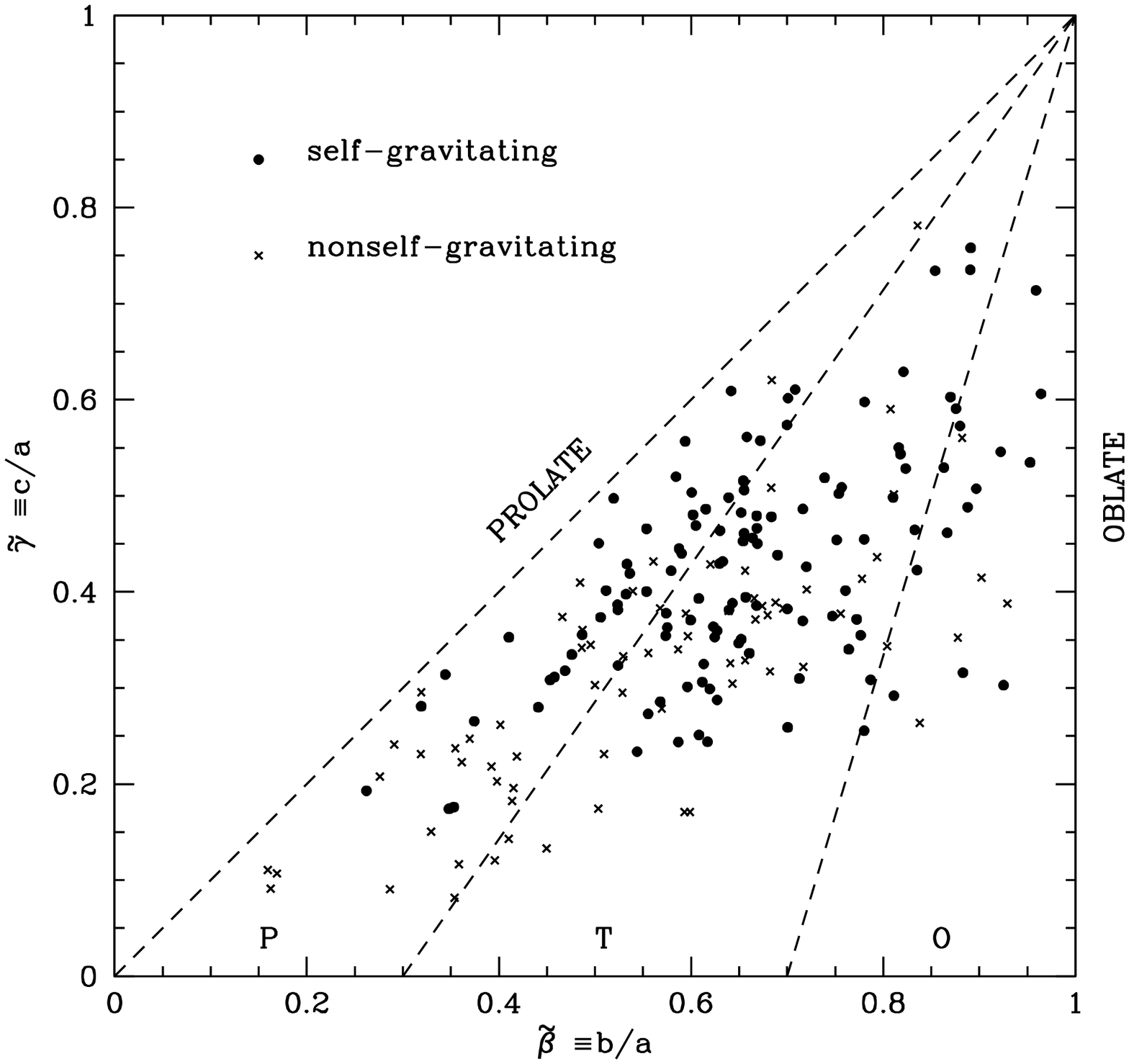}{3.2}{0.5}{-10}{-0}
\figcaption{
The distribution of 3D clump axis ratios in \LC 9.   Clumps lying on the
diagonal are prolate ($b/a = c/a < 1$); clumps lying along the right
hand boundary of the plot are oblate ($a/b = 1$; $c < 1$).  We have
divided the diagram into three regions: prolate (within the triangle
marked $P$); triaxial (within the triangle marked $T$); and oblate
(within the triangle marked $O$).  The precise boundaries of these
regions are arbitrary.
}

\subsection{3D Axis Ratios}

Figure 10 shows the distribution of true axis ratios in \LC9.  The
self-gravitating clumps are denoted as solid dots, while
nonself-gravitating clumps are crosses. The clumps are truly triaxial in
the sense that they do not cluster around the prolate or oblate axes.
As a convenience, we divide the $\tilde{\beta}-\tilde{\gamma}$ plane
into three parts: a ``prolate'' part (which lies above the line
connecting $(\tilde{\beta},\tilde{\gamma})=(1.0,1.0)$ and $(0.33,0.0)$),
an ``oblate'' group (which lies below the line connecting
$(\tilde{\beta},\tilde{\gamma})=(1.0,1.0)$ and $(0.67,0.0)$), and a
``triaxial'' group (everything else).  Of the 300 clumps, 41\% are
prolate, 47\% are triaxial, and only 12\% are oblate.  Although the
relative clump numbers vary from snapshot to snapshot, in general the
prolate and triaxial groups comprise about 90\% of the clumps.  A purely
spheroidal (oblate or prolate) model for the intrinsic clump shape
distribution is not viable.

The prolate, oblate, and triaxial fraction for each snapshot are listed in
Table 3.  The only trend worth remarking on is the growth in prolate clumps
with time in run \LB.  As we shall see below, these objects have a weak
tendency to line up so that the long axis is perpendicular to the magnetic
field.

The broad distribution of clump shapes in the
$\tilde{\beta}-\tilde{\gamma}$ plane is likely telling us that clumps
have complicated internal dynamics-- they are turbulent-- and are not
well described by highly symmetric equilibrium models.  The
self-gravitating clumps, however, are slightly ``rounder'' than
nonself-gravitating ones; they tend to lie closer to the point
$(\tilde{\beta},\tilde{\gamma})=(1.0,1.0)$ than their
nonself-gravitating counterparts (this is not a resolution effect, as
self-gravitating clumps tend to be larger and hence better resolved than
nonself-gravitating clumps).

\begin{table*}[htb]
\begin{center}
\caption{Shape and Magnetic Field Alignment in Clump Snapshots}
\begin{tabular}{lccccccccccc}
\hline
\hline
 & & & & \multicolumn{3}{c}{local alignment}
& \multicolumn{3}{c}{global alignment} & \multicolumn{2}{c}{proj. align.} \\
\cline{5-7} \cline{8-10} \cline{11-12}
Snapshot & \multicolumn{3}{c}{clump shape} & long & int. & short & long & int. & short & long & short \\

no. &  $\mathcal{P}$\tablenotemark{a} & $\mathcal{T}$\tablenotemark{a} & $\mathcal{O}$\tablenotemark{a} & $\mathcal{L.A.}$\tablenotemark{b} &
$\mathcal{L.B.}$\tablenotemark{b} & $\mathcal{L.C.}$\tablenotemark{b} & $\mathcal{G.A.}$ \tablenotemark{c}& $\mathcal{G.B.}$\tablenotemark{c} & $\mathcal{G.C.}$\tablenotemark{c}
&$\mathcal{P.A.}$ \tablenotemark{d} & $\mathcal{P.B.}$ \tablenotemark{d}\\
\hline
\LB 3  & 0.43 & 0.42 & 0.15 & 0.06 & 0.09 & 0.15 & 0.14 & 0.15 & 0.13 & 0.27 & 0.38\\
\LB 7  & 0.48 & 0.41 & 0.11 & 0.11 & 0.13 & 0.14 & 0.13 & 0.12 & 0.13 & 0.30 & 0.36\\
\LB 19 & 0.50 & 0.39 & 0.11 & 0.07 & 0.13 & 0.22 & 0.07 & 0.12 & 0.18 & 0.26 & 0.41\\
\LB 29 & 0.56 & 0.31 & 0.13 & 0.00 & 0.13 & 0.25 & 0.00 & 0.09 & 0.25 & 0.21 & 0.44\\
\hline
\LC 3  & 0.47 & 0.46 & 0.07 & 0.14 & 0.16 & 0.13 & 0.12 & 0.12 & 0.16 & 0.35 & 0.32\\
\LC 4  & 0.46 & 0.42 & 0.12 & 0.14 & 0.14 & 0.12 & 0.14 & 0.11 & 0.16 & 0.34 & 0.32\\
\LC 9  & 0.41 & 0.47 & 0.12 & 0.14 & 0.12 & 0.14 & 0.10 & 0.15 & 0.15 & 0.33 & 0.33\\
\LC 11 & 0.48 & 0.42 & 0.10 & 0.11 & 0.10 & 0.16 & 0.14 & 0.10 & 0.12 & 0.30 & 0.36\\
\LC 19 & 0.41 & 0.47 & 0.12 & 0.10 & 0.12 & 0.20 & 0.14 & 0.12 & 0.15 & 0.29 & 0.39\\
\hline
\LD 3  & 0.42 & 0.44 & 0.14 & 0.21 & 0.17 & 0.04 & 0.10 & 0.12 & 0.11 & 0.44 & 0.24\\
\LD 5  & 0.42 & 0.46 & 0.12 & 0.17 & 0.15 & 0.09 & 0.13 & 0.10 & 0.10 & 0.37 & 0.30\\
\LD 9  & 0.48 & 0.40 & 0.12 & 0.13 & 0.15 & 0.10 & 0.12 & 0.09 & 0.18 & 0.36 & 0.31\\
\LD 11 & 0.42 & 0.45 & 0.13 & 0.14 & 0.14 & 0.12 & 0.10 & 0.13 & 0.14 & 0.36 & 0.31\\
\LD 15 & 0.53 & 0.37 & 0.10 & 0.13 & 0.14 & 0.10 & 0.10 & 0.14 & 0.18 & 0.34 & 0.31\\
\hline
\end{tabular}
\tablenotetext{a}{Fraction of prolate ($\mathcal{P}$), triaxial
($\mathcal{T}$) and oblate ($\mathcal{O}$) clumps.}
\tablenotetext{b}{Fraction of clumps whose density-weighted mean magnetic 
field is within $30\deg$ of the clump major axis ($\mathcal{L.A.}$), 
within $30\deg$ of the intermediate axis ($\mathcal{L.B.}$),
or within $30\deg$ of the major axis ($\mathcal{L.C.}$).
}
\tablenotetext{c}{Fraction of clumps for which the global mean magnetic
field is within $30\deg$ of the clump minor axis ($\mathcal{G.A.}$), etc..}
\tablenotetext{d}{Fraction of clumps for which the projected density
weighted mean magnetic field is within $30\deg$ of
the projected minor axis ($\mathcal{P.A.}$), etc..}
\end{center}
\end{table*}

\subsection{Apparent Axis Ratio}

The procedure for obtaining the apparent axis ratio of a 3D clump was
described in \S \ref{SHAPDEF}.    If we apply this procedure to every
clump and average over many viewing angles, we can obtain a mean, or
expected distribution for $f(r)$.  Figure 8 (solid line) shows the
distribution of the apparent axis ratios obtained from \LC9, as well as
the result from M17SW \citep{sgu90}.

Crudely speaking, the two distributions are consistent in the sense that
both show a characteristic absence of nearly round objects.  It is this
absence that most strongly rules out a parent population of oblate
spheroids \citep{ryd96}.  The $f(r)$ changes with time in the sense that
the mean apparent axis ratio increases.  In fact, $f(r)$ for
snapshots \LC11 and \LC19 lies within the error bounds permitted by the
observations.  A similar increase in the mean apparent axis ratio with
time is seen in run \LD\ (the clumps become rounder with time), but not
run \LB.

Our discussion up to now has concerned apparent axis ratios of
individual clumps.  Clumps exhibit considerable overlap (``confusion''),
however, so the distribution of axis ratios of apparent (2D) clumps may
be different from the distribution of apparent axis ratios of real (3D)
clumps seen in projection.  Figure 9 shows a comparison between the
former (2D clumps obtained from a single line of sight projection) and
the M17 observations.  Broadly speaking, they are consistent with each
other and the 2D clumps also show the absence of round clumps.

\section{Clump Orientation}

Steady-state theoretical models of magnetically supported clouds suggest
that clouds should be flattened perpendicular to the magnetic field
(e.g., \citealt{mou76, tin88a, tin88b, bes89, ls89, gs93, fmou93, ls96})
or, in the language of this paper, that the clump minor axis should be
nearly parallel to the local field.  Observations by \citet{gbmm90} (also
\citealt{agjlm92}) suggest, however, that cloud minor axes are not
correlated with the local field direction.  Recent observations showing
alignment (\citealt{mawi00}) are for a small number of clumps in individual
clouds and probably do not indicate a statistically significant
alignment \footnote{Assessing the statistical significance of alignments
is complicated by the fact that both magnetic fields and clump
orientations are spatially correlated.}.  Here we examine correlations
between the clump principal axes and the {\it density-weighted} mean
field within clumps, as well as the mean global fields, $\< {\bf B} \>$.

\medskip
\myputfigure{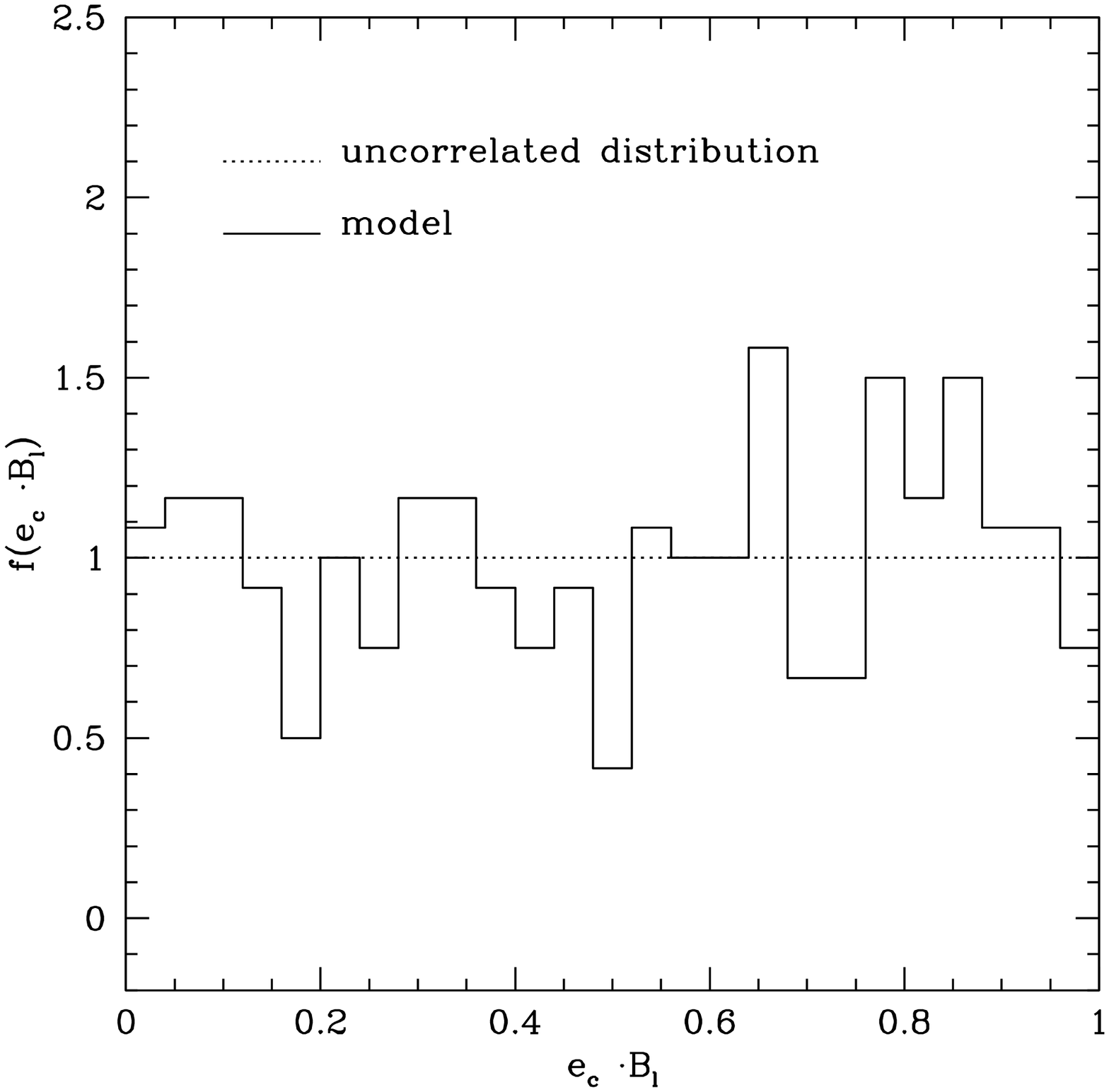}{3.2}{0.5}{-10}{-0}
\figcaption{
The distribution of angles between the shortest body axis and the
density weighted mean magnetic field in the clump in \LC 9.
The dashed line shows the distribution expected if both vectors are
chosen randomly.
}
\medskip

Figure 11 shows the distribution of the dot product of the clump short
axis unit vector (${\bf e}_c$) and the unit vector associated with the
density weighted mean magnetic field in the clump (${\bf b}_l$) in
snapshot \LC9.  The dashed line shows the distribution expected if both
vectors are chosen at random.  Evidently there is no strong correlation
between the mean local field and the clump minor axis.

Figure 12 shows a similar plot for 3D clumps in projection, also for
\LC9.  The distribution is obtained as follows: for a fixed viewing
angle, we use the same procedure described in \S 5 to obtain the
principal axes of the projected clump, as well as the density-weighted
magnetic field vector within each clump. Averaging over $10^4$ viewing
angles gives the distribution. Again there is no
strong correlation between the local mean field and the clump minor
axis.

A more complete description of alignment statistics is given in Table 3.
There we list, for each of the snapshots, the fractional alignment of
clump axes and the global and local mean fields.  The quantity
$\mathcal{L.A.}$, for example, is the fraction of all clumps with $|{\bf
e}_a \cdot {\bf b}_l| > 0.86$, i.e. the fraction of all clumps with
major axis closer than $30^{\circ}$ to the local mean field.  Similarly,
$\mathcal{L.B.}$ is the fraction of all clumps with intermediate axis
closer than $30^{\circ}$ to the local mean field, etc.

Table 3 also describes the alignment of clump axes with the global volume
mean magnetic field with unit vector ${\bf b}_g$.  The quantity
$\mathcal{G.A.}$ is the fraction of all clumps with $|{\bf e}_a \cdot
{\bf b}_g| > 0.86$.  The only significant global alignment is in the
late stages of the strong field run, when the short axis has a weak
tendency to line up with the global mean field.

\medskip
\myputfigure{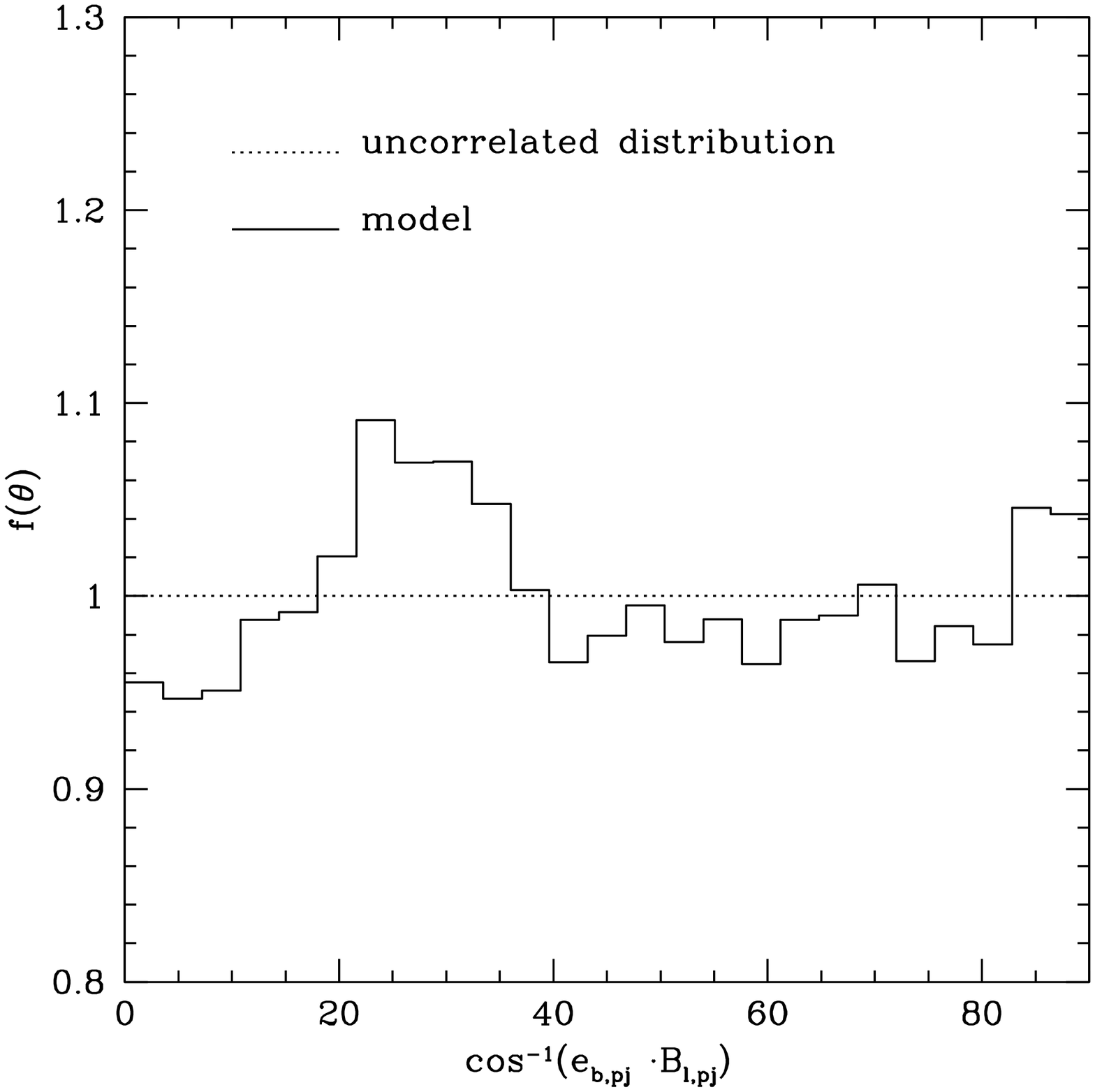}{3.2}{0.5}{-10}{-0}
\figcaption{
The distribution of angles between the projected shortest body axis and
the projected density weighted mean magnetic field in \LC 9.
The dashed line shows the distribution expected if both vectors are
chosen randomly.
}
\medskip

Notice that if clump magnetic fields are randomly oriented with respect
to the clump principal axes, {\it and} if the orientation of each clump
is independent, we expect $\mathcal{L.A.} = 0.14 \pm 0.52/\sqrt{N}$ at
the $87\%$ confidence level, where $N$ is the number of clumps (and
similarly for $\mathcal{L.B.}, \mathcal{L.C.}$, etc.).  This na\"{\i}ve
estimate for the dispersion in alignments is not accurate because clump
orientations are probably spatially correlated (i.e. not independent).
For example consider run \LD, whose weak initial field makes any global
alignments physically implausible.  Nevertheless, \LD\ exhibits global
alignments that are ``statistically significant'' according to our
na\"{\i}ve estimate.

The only significant and physically plausible alignments occur in run
\LB, where we find that the minor axis tends to be weakly aligned with
the field. The longest axis also tends to be anti-aligned with the
field.  This is similar to what might be expected from an equilibrium
model, except that the clump shapes tend to be dominantly prolate or
triaxial rather than oblate.

Table 3 also describes the alignment of projected mean field vectors
with the principal axes of the projected moment of inertia tensor.
These alignments are given in the final two columns ($\mathcal{P.A.}$
and $\mathcal{P.B.}$, which are the fraction of clumps with projected
field less than $30^{\circ}$ from the projected clump major and minor
axes, respectively). For randomly oriented clumps we expect
$\mathcal{P.A.} = \mathcal{P.B.} = 0.33$. The projected alignments are
sensitive to viewing angle, so the values shown have been averaged over
$10^4$ uniformly distributed viewing angles.  The only plausible
alignments are in run \LB; other snapshots show at best marginally
significant alignments.  To sum up, our numerical models suggest that
projected alignments are weak and not a useful diagnostic of field
strength.

Clump principal axis alignments, like the mass spectrum, are sensitive
to the smoothing parameter $\Delta$.  For example, if $\Delta = 0$ we
see an implausible set of alignments in run \LD.  Presumably this is
because we are studying alignments near the grid scale.  This is one of
the motivations for introducing a finite $\Delta$.

\section{Rotation}

Observations suggest that rotation is not important in supporting clumps
against gravitational collapse since rotational energy is small compared
to the gravitational potential energy (e.g. \citealt{gbfm93}).  However,
once collapse takes place, angular momentum may play an important role
in determining the characteristics of the system to be formed, e.g., the
size of circumstellar disks, or the separation of the stars in a
multiple star system.

Clumps or cores are observed to have greater specific angular momentum
$j$ than binary or multiple star systems.  Consider a binary with masses
of $1 \msun$ for both members and a period of $100$ years. The specific
angular momentum of this system is $j_{bin} =  1.16 \times 10^{20}\,{\rm
cm^2\,s^{-1}}$.  This is typical for young binary T Tauri stars (e.g.
\citealt{setal95}). For single protostars and the envelope, \citet{oha97}
obtained $j \sim 10^{20} \,{\rm cm^2\,s^{-1}}$ for two IRAS sources (for
circumstellar disks, see, e.g. \citealt{beck90}).  As for dense cores, the
results given by \citet{gbfm93} range from $6 \times 10^{20}$ to
$4 \times 10^{22}\,{\rm cm^2\,s^{-1}}$.

\medskip
\myputfigure{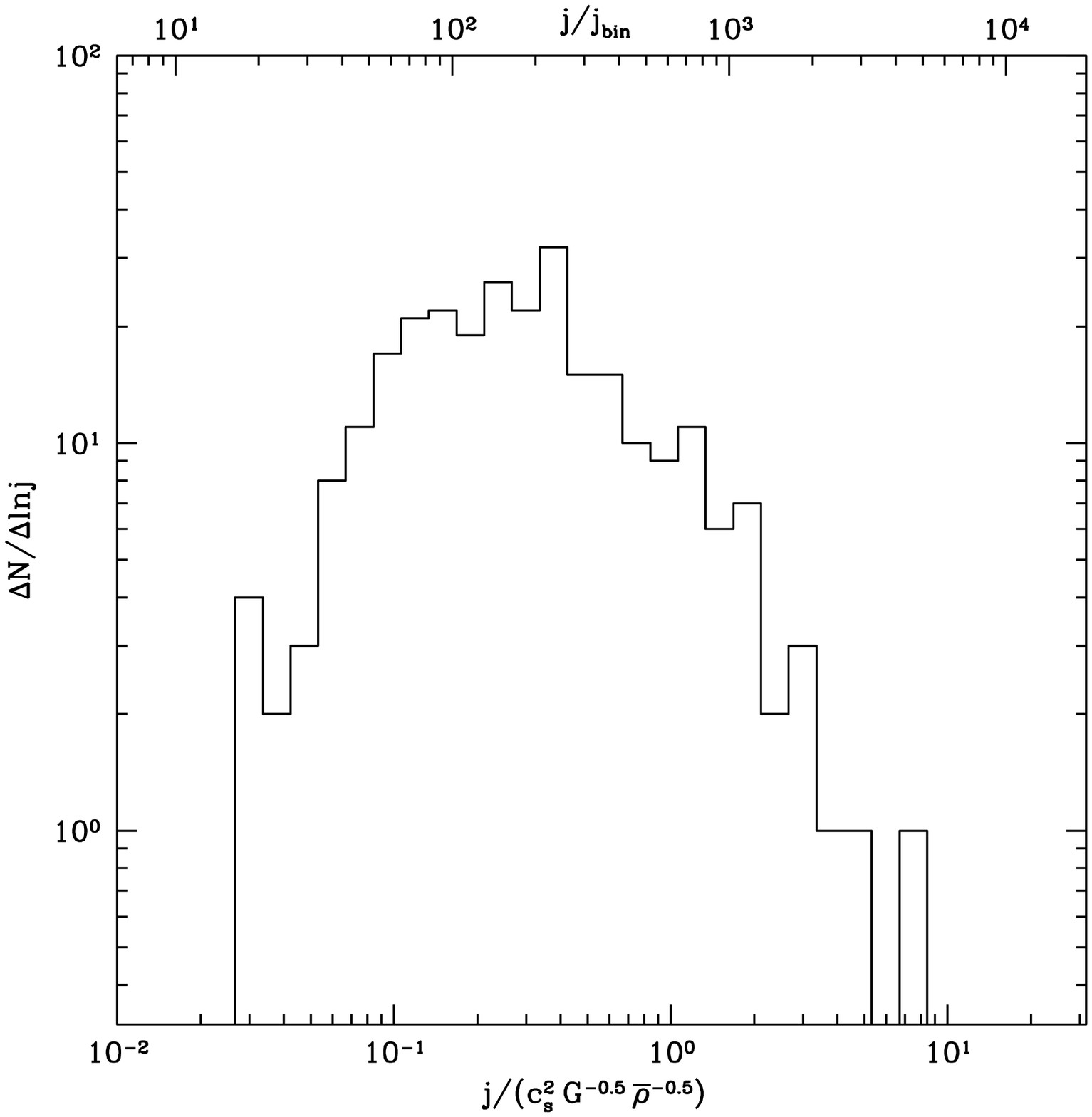}{3.2}{0.5}{-10}{-0}
\figcaption{
Distribution of specific clump spin angular momentum in \LC 9. The
normalization is to $c^2_s\,G^{-0.5}\,\bar{\rho}^{-0.5} \equiv
j_{tot}/(n_J\,\pi^{0.5})$, where $j_{tot} \equiv 3.8\times 10^{23}\,{\rm
cm^2\,s^{-1}} = 3280\,j_{bin}$ and $j_{bin} \approx 10^{20} {\rm
cm^2\,s^{-1}}$ is a reference angular momentum for a pair of $1\msun$
stars in circular orbits at $100\au$.
}
\medskip

We have examined the specific angular momenta of the simulated clumps,
and find that the values are greater than $j_{bin}$. In Figure 13 we
show the distribution of specific angular momenta of clumps in \LC9.
Clump specific angular momenta span about $2-3$ decades, e.g., from
$10\,j_{bin}$ to $\sim 6\times 10^{3}\,j_{bin}$, the peak is at
$2.5\times 10^2\,j_{bin}$; the range of the distribution decreases with
time in \LC\ and \LD\ snapshots (we cannot say much about run \LB\
because there are not many clumps in later snapshots).  There is no
clear trend in clump specific angular momentum with mass.

Our distribution of specific angular momenta may be compared to those
found by \citet{bbd00} (hereafter BB), who studied clump models in which
the internal velocities are drawn from a Gaussian random field at a
spatial scale $\sim 0.1 {\rm pc}$; their power spectrum ($v^2 \propto
k^{-4}$) was normalized to match the typical velocity dispersion in the
\citet{goo98} sample of cores.  Their distribution peaks at $\sim
2\times 10^{21} \cm^2 \sec^{-1}$, while ours peaks at $4\times 10^{22}
\cm^2 \sec^{-1}$.  This difference is attributable to the following
factors.  First, our sample contains both nonself-gravitating and
self-gravitating clumps, while BB's clumps are scaled to match
self-gravitating clumps.  Second, the self-gravitating clumps in our
sample are more massive (by a factor of about $4$) than BB's using our
standard scaling.  Third, the self-gravitating clumps have lower mean
specific angular momentum (by about $40\%$) than the nonself-gravitating
clumps.  Fourth, while BB chose a Gaussian clump density profile, our
clumps have substantial envelopes that can make a significant
contribution to clump angular momentum.  We conclude that the difference
is not significant.

\section{Summary}

The main results of this paper are:

1. Clumps are triaxial.  Clumps that are ``prolate'' or ``triaxial'' (in
the sense defined in \S 5) make up $90\%$ of the clump population.
Clumps that are ``oblate'' make up only $10\%$ of the clump population.
The distribution of clump shapes implies a distribution of apparent axis
ratios that is consistent with observations.  We do not see any signs of
the sort of oblate equilibria that have been the focus of earlier
studies of magnetized clouds emerging in our simulations.  But we have
not considered the most general possible set of initial states (we
assume the mass-to-flux ratio is uniform), nor have we included the
effects of ambipolar diffusion.

The possibility of triaxial cloud cores has also been explored by
\citet{b00,jbd01,jb02}.  \citet{b00}, however, suggested that the short
axis should be aligned with the magnetic field.  This is not seen in
our numerical models.  The absence of magnetically supported equilibria
in self-gravitating, turbulent MHD models has also been noted by several
other workers in the field, perhaps most forcefully by \citet{pnrg00}.
The emergence of such equilibria in supersonic, self-gravitating
turbulence cannot yet be ruled out, however, because the effects of
ambipolar diffusion, varying mass to flux ratios, and varying initial
conditions have not yet been studied.

2.  Clump principal axes are not strongly aligned with the magnetic
field, either locally or globally.  We would not expect to see such
alignments in observations, nor are they currently observed.  The effect
of magnetic fields on cloud structure has been considered earlier by
\citet{bpm02}.  They found that only strong fields produce an observable
effect on the density structure based on a qualitative analysis.  A
similar result was seen at late stages of self-gravitating evolution in
the strong-field models of \citet{ogs99}.

3. The simulated clump mass spectrum has a characteristic mass of $\sim
0.5 \msun$ (for an assumed mean cloud density and temperature).  While
this result is suggestive, it depends on some of the choices made in
setting up our simulations, including numerical resolution.   It is
possible that the characteristic mass scales with the total cloud mass
or with the sonic Mach number or density of the initial conditions.
These scalings have not yet been explored.  The high-mass wing of the
clump mass spectrum has a slope that is crudely consistent with the
Salpeter law, although other, non-power-law forms for the mass spectrum
may well be consistent with our data.

Other numerical work on the clump mass spectrum by
\citet{pnrg00,kl01,kb01,pn02,bpm02} is generally consistent with our
results.  These models include similar physics to ours, although some
models lack selfgravity \citep{bpm02} , while others lack magnetic
fields \citep{kl01,kb01}.

The simulations of \citet{pnrg00} are the most similar to those analyzed
here.   Differences in Padoan et al.'s analytical methods limit the
possibility of detailed comparisons, however:  their definition of a
self-gravitating clump does not include turbulent kinetic energy; the
mass spectra presented are composites of two models with gas
temperatures that differ by a factor of three; their definition of a
clump depends on contrast of local peaks rather than a density
threshold.  Nevertheless, the results of the two analyses for the
position of the peak of the mass spectrum, as well as for the slope of
the high-mass end, are broadly consistent.

\citet{kl01} and \citet{kb01} use a completely different numerical method
(smoothed particle hydrodynamics) and include selfgravity but not magnetic 
fields.  In addition,
they allow dense regions to turn into ``sink'' particles.  This allows
them to continue the evolution much further in time than is possible in
our calculations.  The trends we find of an increasingly flat high-end
spectrum, and a larger proportion of mass in bound objects, are clearly
evident in their distributions at sequential evolutionary stages.  Their
clump mass spectra are again crudely consistent with ours; minor
differences in mass scale and shape may arise due to differences in
initial conditions or input parameters (number of Jeans masses).  We
cannot evaluate this without a direct attempt to reproduce their models.

Recently, \citet{pn00} (see also \citealt{pn02}) have proposed that the
clump mass spectrum, and consequently the mass function of collapsing
cores, may be predicted solely from three macroscopic quantities
averaged over the entire cloud: density, temperature, and velocity
dispersion.  Although this theory is attractive in many way, the
stationary turbulent state adopted for the \citet{pn00} may not be a good
approximation for real star forming regions.  Clouds are transient
structures, and a cloud may have developed its current mean density,
temperature, and velocity dispersion in many ways.  Our analysis of
purely decaying models shows that some aspects of the mass spectrum are
invariant in time (e.g.  $M_{peak}$) even as the Mach number decreases,
while other aspects of the mass spectrum (the fraction of
self-gravitating clumps) increase secularly with time.  Both of these
results suggest that a cloud retains a memory of its dynamical history
that may be important for establishing the stellar IMF.

4. Clump specific angular momenta are an order of magnitude larger than
in a ``typical'' binary system.  It is interesting that our simulations
give clump angular momenta that are so close to the angular momenta of
binary systems.  This suggests that we are close to resolving the
lengthscale where the angular momentum of binaries is determined (since
$\Delta v_\lambda \sim \lambda^{1/2}$ in molecular clouds and in
supersonic turbulence generally, the specific angular momentum $j \sim
\Delta v_\lambda \lambda \sim \lambda^{3/2}$, so $\lambda \sim
j^{2/3}$).  Future higher resolution numerical models, perhaps with an
adaptive mesh, will resolve this scale and may be capable of directly
simulating binary formation.  

Our results all depend on how the clumps are defined.  Because
fragmentation during gravitational collapse may decide the final stellar
mass function, characterizing the hierarchy of clumpy structure and
understanding its consequences may be very important.  Mass functions
are steeper when clumps-within-clumps are counted rather than discounted
(see e.g. \citealt{ost02}); whether gravity eventually selects the larger
or smaller scale of a nested structure depends on the details of the
collapse.  More general tools (e.g.  wavelet methods) for analyzing the
density field will be valuable in characterizing this hierarchical
structure.  Future analyses should also focus on direct analogues of
observable clumps (e.g. simulated clumps in position-position-velocity
space for common molecular tracers), although self-consistent treatments
of thermodynamics, radiative transfer, and molecular chemistry may
ultimately be necessary for realistic source functions.

This work was supported by NASA grants NAGW 5-9180 and NAG 59167.

\end{document}